\documentclass[10pt,aps,pra,onecolumn]{revtex4-2}
\usepackage{amsmath}
\usepackage{amssymb}
\usepackage{graphicx}
\usepackage{hyperref}
\usepackage[english]{babel}
\usepackage{babel}
\usepackage[usenames,dvipsnames]{color}

\newcommand{\bear}{\begin{eqnarray}}
\newcommand{\eear}{\end{eqnarray}}
\newcommand{\be}{\begin{equation}}
\newcommand{\ee}{\end{equation}}
\newcommand{\beqn}{\begin{eqnarray}}
\newcommand{\eeqn}{\end{eqnarray}}
\newcommand{\beqnn}{\begin{eqnarray*}}
\newcommand{\eeqnn}{\end{eqnarray*}}
\input{tcilatex}
\begin{document}

\title{Comparison of Quantum and Semiclassical Rabi models near multiphoton resonances in the presence of parametric modulation}
\author{M V S de Paula$^{1}$, M A Damasceno Faustino$^{1}$ and A V Dodonov$^{1,2}$}
\email{adodonov@unb.br}
\affiliation{$^1$Instituto de F\'{\i}sica, Universidade de Bras\'{\i}lia, Caixa Postal 04455, CEP
70910-900, Bras\'{\i}lia, DF, Brasil \\
$^2$International Center of Physics, Institute of Physics, University of Brasilia, 70910-900, Brasilia, DF, Brazil}

\begin{abstract}
We compare the semiclassical and quantum predictions for the unitary dynamics of a two-level atom interacting with a single-mode electromagnetic field in the presence of parametric modulation of the atomic parameters, in the regime of multiphoton atom-field resonances. We derive approximate analytic solutions for the semiclassical Rabi model when the atomic transition frequency and the atom-field coupling strength undergo harmonic external modulations. These solutions are compared to the predictions of the quantum Rabi model, which is solved numerically for the initial coherent state with a large average photon number, of the order of $10^4$, in the regime of three-photon resonance. We show that for initial times and sufficiently intense coherent state the semiclassical dynamics agrees quite well with the quantum one, although for large times it inevitably fails due to the lack of the collapse-revival behavior. Furthermore, we describe how the field state is modified throughout the evolution, presenting numeric results for the average photon number, entropies (related to the atom-field entanglement) and other quantities that characterize the photon number statistics of the electromagnetic field.
\end{abstract}

\maketitle

\section{Introduction}

The semiclassical Rabi model (SRM) \cite{rabi1,rabi2,bloch} plays a
fundamental role in understanding the interaction between light and matter,
particularly in regimes where the electromagnetic field can be treated
classically while the atomic system retains its quantum nature \cite{irish}.
It provides essential insights into the dynamics of two-level systems
(qubits) driven by external fields, capturing key phenomena such as Rabi
oscillations, parametric modulations, single- and multiphoton resonances,
etc \cite{shirley,duvall,woerd,ad7,castanos,saiko,ad3,simse,marinho0,marinho1,marinho2}.
Despite its apparent simplicity, it forms the basis for experimental
implementations in areas ranging from laser--atom interactions to quantum
control protocols \cite{boyd,scully,shore}. Since the SRM ignores the quantum nature of
light, it fails to accurately describe the dynamics of the qubit for all
times, besides overlooking the correct physical interpretation of the qubit excitation
as a result of the absorption of photons \cite{luo,acosta1}. These limitations are overcome by the quantum Rabi model (QRM) \cite{braak,rev,larson}, in
which the quantum nature of the light is fully taken into account, and
resonances and parametric modulations are explained from the quantum point
of view \cite{klim0,j1,j3,j2,3fot1,garzi,j4,ma,3fot3,luo}.

In this paper, our first goal is to study the SRM in the presence of
parametric modulation of system parameters, assuming the regime of multiphoton resonances. From a physical point of view, this means investigating how
complete oscillations of the qubit excitation probability in the far
dispersive regime emerge when the qubit transition frequency and the
frequencies of external modulations are properly adjusted. In particular, we
consider separate or simultaneous modulations of the qubit transition
frequency or the qubit--field coupling parameter, and analyze in detail the first-
and second-order effects with respect to the modulation amplitudes. Our
second goal is to give a full characterization of the qubit behavior according to the QRM,
pinpointing the differences between the semiclassical and the quantum
predictions for the qubit excitation probability. Lastly, we investigate how
the excitation and deexcitation of the qubit affect the field state of the
electromagnetic field. For that, we assume that the cavity field is prepared
in a coherent state with a large amplitude, and study the behavior of the
average photon number, von Neumann and linear entropies, and some other
quantities related to the photon number statistics and deviation of the field from the initial state. We show
that for initial times and sufficiently intense coherent states, the
semiclassical and quantum results coincide, but for larger times,
quantitative and qualitative discrepancies appear. Moreover, our numeric
simulations provide concrete estimates of the qubit's purity during the
evolution, which is related to the degree of the atom--field entanglement (in the current scenario of a bipartite lossless system).

This paper is organized as follows. In Section \ref{semiclas} we present the
approximate analytic solution of the SRM in the presence of external
modulations, when the conditions for a multiphoton resonance are met. In Section \ref%
{quant} we consider the QRM at the three-photon resonance condition, and
analytically show how the resonant external modulations can affect the
system dynamics for the initial coherent state with a large average photon
number. In Section \ref{numeric} we present the numeric results for the
system dynamics according to the QRM, considering initial coherent
states with the average photon numbers equal to $5\times 10^{3}$ and $%
3\times 10^{4}$, and compare the predictions of SRM and QRM. Finally,
Section \ref{conc} contains our conclusions. In Appendix \ref{appen} we
present closed analytic expressions for some transition rates relevant to
this work.

\section{Semiclassical Rabi model\label{semiclas}}

Let us consider the semiclassical Rabi Hamiltonian \cite{rabi1,rabi2}  with
externally prescribed qubit transition frequency $\Omega (t)=\Omega
_{0}+\varepsilon _{\Omega }\sin \left( \eta _{\Omega }t\right) $ and the
atom--field coupling strength $g(t)=g_{0}+\varepsilon _{g}\sin ( \eta
_{g}t) $
\begin{equation}
\hat{H}_{S}(t)=\frac{\Omega (t)}{2}\hat{\sigma}_{z}+g(t)\left(
E_{1}+E_{-1}\right) \hat{\sigma}_{x}~,  \label{zer}
\end{equation}%
where $\hat{\sigma}_{i}$ are the Pauli matrices and we set $\hbar=1$. We use the short-hand
notation $E_{k+X}\equiv \exp \left[ it\left( k\nu +X\right) \right] $, where $\nu$ is the field frequency,
$t$ is the time, $k$ is an integer and $X$ is an arbitrary real parameter.
In the interaction picture defined by the operator $\hat{U}_{1}=\exp \left(
-i\nu t\hat{\sigma}_{z}/2\right) $, the Hamiltonian reads%
\begin{equation}
\hat{H}_{I}(t)=\hat{H}_{ef}+\hat{H}_{t}(t)\,,  \label{H1}
\end{equation}%
\begin{equation}
\hat{H}_{ef}=\frac{\Omega _{0}-\nu }{2}\hat{\sigma}_{z}+g_{0}\hat{\sigma}_{x}
\end{equation}%
\begin{equation}
\hat{H}_{t}(t)=\frac{\epsilon _{z}(t)}{2}\hat{\sigma}_{z}+\epsilon _{+}(t)%
\hat{\sigma}_{+}+\epsilon _{-}(t)\hat{\sigma}_{-}~,
\end{equation}%
where $\hat{\sigma}_{+}=\hat{\sigma}_{-}^{\dagger }=|e\rangle \langle g|$ is
the ladder operator, $|g\rangle $ and $|e\rangle $ denote the ground and the
excited states of the atom, and we defined the time-dependent functions $%
\epsilon _{z}(t)=\varepsilon _{\Omega }\sin \left( \eta _{\Omega }t\right) $%
, $\epsilon _{+}(t)=\epsilon _{-}^{\ast }(t)=\varepsilon _{g}\sin \eta
_{g}t+g(t)e^{2i\nu t}$.

The time-independent part of the interaction-picture Hamiltonian, denoted as
$\hat{H}_{ef}$ in Eq. (\ref{H1}), has the eigenenergies $\pm R/2$ with the
corresponding eigenstates%
\begin{equation}
|\phi _{+}\rangle =\tilde{R}_{+}^{-1/2}\left( \tilde{R}_{+}|g\rangle +\tilde{%
g}_{0}|e\rangle \right) ~,~|\phi _{-}\rangle =\tilde{R}_{-}^{-1/2}\left(
\tilde{R}_{-}|g\rangle -\tilde{g}_{0}|e\rangle \right) ~,
\end{equation}%
where $R=\sqrt{4g_{0}^{2}+(\nu -\Omega _{0})^{2}}$ and we defined $\tilde{%
\Delta}=(\nu -\Omega _{0})/R$, $\tilde{g}_{0}=g_{0}/R$ and $\tilde{R}_{\pm
}=\left( 1\pm \tilde{\Delta}\right) /2$.

\subsection{Jacobi-Anger expansion}

To obtain a closed approximate description of the system dynamics, we expand
the wavefunction corresponding to the Hamiltonian $\hat{H}_{I}(t)$ as

\begin{equation}
|\psi _{1}\rangle =A_{+,1}(t)e^{iT_{2}/2}e^{-itR/2}|\phi _{+}\rangle
+A_{-,1}(t)e^{-iT_{2}/2}e^{itR/2}|\phi _{-}\rangle\,,
\end{equation}%
where%
\begin{eqnarray}
T_{2} &=&\varphi -\Upsilon _{1}\cos \left( \eta _{\Omega }t\right) -\Upsilon
_{0}\cos \left( 2\nu t-\pi /2\right) \\
&&+\Upsilon _{2}\cos \left( \eta _{g}t\right) +\Upsilon _{3}\cos \left[
\left( \eta _{g}+2\nu \right) t\right] +\Upsilon _{4}\cos \left[ \left( \eta
_{g}-2\nu \right) t\right]  \nonumber
\end{eqnarray}%
\begin{equation}
\Upsilon _{0}=g_{0}\frac{2\tilde{g}_{0}}{\nu }~,~\Upsilon _{1}=\varepsilon
_{\Omega }\frac{\tilde{\Delta}}{\eta _{\Omega }}~,~\Upsilon _{2}=\varepsilon
_{g}\frac{4\tilde{g}_{0}}{\eta _{g}}
\end{equation}%
\begin{equation}
\Upsilon _{3}=\varepsilon _{g}\frac{2\tilde{g}_{0}}{\eta _{g}+2\nu }%
~,~\Upsilon _{4}=\varepsilon _{g}\frac{2\tilde{g}_{0}}{\eta _{g}-2\nu }
\end{equation}%
and $\varphi =\Upsilon _{1}-\Upsilon _{2}-\Upsilon _{3}-\Upsilon _{4}\,$.
For the common initial condition, $|\psi _{1}\left( 0\right) \rangle
=|g\rangle $, one has $A_{+,1}(0)=\tilde{R}_{+}^{1/2}$ and $A_{-,1}(0)=\tilde{%
R}_{-}^{1/2}$.

The exact system dynamics is obtained by solving the pair of equations%
\begin{equation}
\dot{A}_{-,1}=-iQ(t)A_{+,1}~,~\dot{A}_{+,1}=-iQ(t)^{\ast }A_{-,1}  \label{e1}
\end{equation}%
with the time-dependent function%
\begin{equation}
Q(t)=e^{-itR}e^{iT_{2}}\left[ \tilde{R}_{-}\epsilon _{+}^{\ast }(t)-\tilde{g}%
_{0}\epsilon _{z}(t)-\tilde{R}_{+}\epsilon _{+}(t)\right] \,. \label{qt}
\end{equation}%
Using the Jacobi-Anger expansion%
\begin{equation}
e^{iz\cos \theta }=J_{0}\left( z\right) +\sum_{n=1}^{\infty
}i^{n}J_{n}\left( z\right) \left( e^{i\theta n}+e^{-i\theta n}\right)
\label{expan}
\end{equation}%
one can recast $Q(t)$ as%
\begin{equation}
Q(t)=\sum_{j=1}^{\infty }p_{j}e^{itf_{j}}\equiv Q_{f}(t)+Q_{s}(t)\,,
\end{equation}%
where $Q_{s}(t)$ denotes the slowly-varying terms (for which $%
|f_{j}|\lesssim |p_{j}|$) and $Q_{f}(t)$ -- the rapidly-varying terms (for
which $\left\vert f_{i}\right\vert \gg \left\vert p_{i}\right\vert $). All
the constant coefficients $p_{j}$ and frequencies $f_{j}$ can be obtained
after straightforward but lengthy manipulations.

So far, all the calculations are exact. However, under the experimentally
feasible conditions $|\Upsilon _{k}|\ll 1~\forall k$, we can expand $Q(t)$
in orders of $\varepsilon _{g}$ and $\varepsilon _{\Omega }$; to abbreviate
the notation, by \textquotedblleft $k$-th order terms in $\varepsilon $%
\textquotedblright\ we shall mean the terms proportional to $\varepsilon
_{g}^{k}$, $\varepsilon _{\Omega }^{k}$ and $\varepsilon _{g}^{n}\varepsilon
_{\Omega }^{{k-n}}$. It is advantageous to first consider the
rapidly-varying terms $Q_{f}(t)$, which to the 0th-order in $\varepsilon $
read%
\begin{equation}
Q_{f}^{(0)} =\Xi _{-R}E_{-R}+\Xi _{-2-R}E_{-2-R}+\Xi _{-4-R}E_{-4-R}
+\Xi _{-6-R}E_{-6-R}+\Xi _{-8-R}E_{-8-R}+\cdots\,,
\end{equation}%
where the time-independent coefficients are%
\begin{equation}
\Xi _{-R}=-g_{0}e^{i\phi }J_{1}^{0}J_{0}^{1}J_{0}^{2}J_{0}^{3}J_{0}^{4}
\end{equation}%
\begin{equation}
\Xi _{-2-R}=e^{i\phi }g_{0}J_{0}^{1}J_{0}^{2}J_{0}^{3}J_{0}^{4}\left(
J_{0}^{0}\tilde{R}_{-}-J_{2}^{0}\tilde{R}_{+}\right)
\end{equation}%
\begin{equation}
\Xi _{-4-R}=e^{i\phi }g_{0}J_{0}^{1}J_{0}^{2}J_{0}^{3}J_{0}^{4}\left(
J_{1}^{0}\tilde{R}_{-}-J_{3}^{0}\tilde{R}_{+}\right)
\end{equation}%
\begin{equation}
\Xi _{-6-R}=e^{i\phi }g_{0}\tilde{R}%
_{-}J_{2}^{0}J_{0}^{1}J_{0}^{2}J_{0}^{3}J_{0}^{4}
\end{equation}%
\begin{equation}
\Xi _{-8-R}=e^{i\phi }g_{0}\tilde{R}%
_{-}J_{3}^{0}J_{0}^{1}J_{0}^{2}J_{0}^{3}J_{0}^{4}
\end{equation}%
and we defined $J_{i}^{k}\equiv J_{i}\left( \Upsilon _{k}\right) $. In
Table \ref{table1} we show the values of above coefficients for $%
g_{0}=0.1\nu $. We see that, at least for $\Omega _{0}>1.5\nu $, $\left\vert
\Xi _{-2-R}\right\vert $ is at least two orders of magnitude larger than the
other coefficients.

\begin{table}[tbp]
\begin{tabular}{|l||l|l|l|l|l|}
\hline
Coefficient & $\Omega _{0}=1.5\nu $ & $\Omega _{0}=2\nu $ & $\Omega
_{0}=2.5\nu $ & $\Omega _{0}=3\nu $ & $\Omega _{0}=3.5\nu $ \\ \hline\hline
$\left\vert \Xi _{-R}\right\vert $ & $1.9\times 10^{-3}$ & $9.8\times
10^{-4} $ & $6.6\times 10^{-4}$ & $5\times 10^{-4}$ & $4\times 10^{-4}$ \\
\hline
$\left\vert \Xi _{-2-R}\right\vert $ & $9.6\times 10^{-2}$ & $9.9\times
10^{-2}$ & $10^{-1}$ & $10^{-1}$ & $10^{-1}$ \\ \hline
$\left\vert \Xi _{-4-R}\right\vert $ & $1.8\times 10^{-3}$ & $9.7\times
10^{-4}$ & $6.6\times 10^{-4}$ & $5\times 10^{-4}$ & $4\times 10^{-4}$ \\
\hline
$\left\vert \Xi _{-6-R}\right\vert $ & $1.7\times 10^{-5}$ & $4.8\times
10^{-6}$ & $2.2\times 10^{-6}$ & $1.2\times 10^{-6}$ & $7.9\times 10^{-7}$
\\ \hline
$\left\vert \Xi _{-8-R}\right\vert $ & $10^{-7}$ & $1.6\times 10^{-8}$ & $%
4.8\times 10^{-9}$ & $2\times 10^{-9}$ & $1.1\times 10^{-9}$ \\ \hline
\end{tabular}%
\caption{Some values of the 0th-order coefficients for different values of $%
\Omega _{0}$ and $g_{0}=0.1\protect\nu $.}
\label{table1}
\end{table}

\subsection{Coarse-grained approximation}

Let us solve Eqs. (\ref{e1}) assuming $Q_{s}=0$ and $Q_{f}(t)=%
\sum_{i=1}^{K}p_{i}e^{itf_{i}}$, where $K$ is some integer, $\left\vert
p_{i}\right\vert \ll \left\vert f_{i}\right\vert $ and $\left\vert
p_{j}\right\vert \ll \left\vert p_{1}\right\vert ~\forall j>1$ (as occurs in
the present work for $\Omega _{0}>1.5\nu $, where $p_{1}=\Xi _{-2-R}$ and $%
f_{1}=-2\nu -R$). One obtains the second-order differential equations%
\begin{equation}
\ddot{A}_{-,1}-\frac{\dot{Q}_{f}}{Q_{f}}\dot{A}_{-,1}+\left\vert
Q_{f}\right\vert ^{2}A_{-,1}=0  \label{AP}
\end{equation}%
\begin{equation}
\ddot{A}_{+,1}-\left( \frac{\dot{Q}_{f}}{Q_{f}}\right) ^{\ast }\dot{A}%
_{+,1}+\left\vert Q_{f}\right\vert ^{2}A_{+,1}=0\,,  \label{AM}
\end{equation}%
where%
\begin{equation}
\frac{\dot{Q}_{f}}{Q_{f}}\approx i\left[ f_{1}+\sum_{j=2}^{K}\left(
f_{j}-f_{1}\right) \frac{p_{j}}{p_{1}}e^{it\left( f_{j}-f_{1}\right)
}-\sum_{j,k=2}^{K}f_{k}\frac{p_{k}}{p_{1}}\frac{p_{j}}{p_{1}}e^{it\left(
f_{k}+f_{j}-2f_{1}\right) }\right]
\end{equation}%
\begin{equation}
\left\vert Q_{f}\right\vert ^{2}=\sum_{j,k=1}^{K}p_{j}p_{k}^{\ast
}e^{it\left( f_{j}-f_{k}\right) }\,.
\end{equation}%

Under the conditions $\left\vert f_{j}+f_{k}-2f_{1}\right\vert \gtrsim
\left\vert f_{j}\right\vert ,\left\vert f_{k}\right\vert $ for $j,k>1$ and $%
\left\vert f_{j}-f_{k\neq j}\right\vert \gtrsim \left\vert f_{1}\right\vert
~\forall j$ we can make the \emph{coarse-grained approximation} by assuming
that $A_{\pm ,1}$ change slowly on the time scale $\tau =2\pi /\left\vert
f_{1}\right\vert $ and averaging the dynamics over $\tau $. We get%
\begin{equation}
\ddot{A}_{-,1}-\dot{A}_{-,1}\frac{1}{\tau }\int_{t}^{t+\tau }dt^{\prime }%
\frac{\dot{Q}_{f}(t^{\prime })}{Q_{f}(t^{\prime })}+A_{-,1}\frac{1}{\tau }%
\int_{t}^{t+\tau }dt^{\prime }\left\vert Q_{f}(t^{\prime })\right\vert
^{2}\approx 0
\end{equation}%
and under our assumptions%
\begin{eqnarray}
\frac{1}{\tau }\int_{t}^{t+\tau }dt^{\prime }\frac{\dot{Q}_{f}(t^{\prime })}{%
Q_{f}(t^{\prime })} &\approx &if_{1}+\frac{\left\vert f_{1}\right\vert }{%
2\pi }\sum_{j=2}^{K}\frac{p_{j}}{p_{1}}e^{it\left( f_{j}-f_{1}\right)
}\left( e^{i\tau f_{j}}-1\right)  \nonumber \\
&&-\frac{\left\vert f_{1}\right\vert }{2\pi }\sum_{j,k=2}^{K}\frac{f_{k}}{%
f_{k}+f_{j}-2f_{1}}\frac{p_{k}p_{j}}{p_{1}^{2}}e^{it\left(
f_{k}+f_{j}-2f_{1}\right) }\left( e^{i\tau \left( f_{k}+f_{j}\right)
}-1\right)  \nonumber \\
&\approx &if_{1}
\end{eqnarray}%
\begin{eqnarray}
\frac{1}{\tau }\int_{t}^{t+\tau }dt^{\prime }\left\vert Q_{f}(t^{\prime
})\right\vert ^{2} &=&\sum_{j=1}^{K}\left\vert p_{j}\right\vert ^{2}-i\frac{%
\left\vert f_{1}\right\vert }{2\pi }\sum_{j\neq k}^{K}\frac{p_{j}p_{k}^{\ast
}}{f_{j}-f_{k}}e^{it\left( f_{j}-f_{k}\right) }\left( e^{i\tau \left(
f_{j}-f_{k}\right) }-1\right)  \nonumber \\
&\approx &\sum_{j=1}^{K}\left\vert p_{j}\right\vert ^{2}\,.
\end{eqnarray}

Thus, under the coarse-grained approximation, Eqs. (\ref{AP}) -- (\ref{AM})
become%
\begin{equation}
\ddot{A}_{\pm ,1}\pm if_{1}\dot{A}_{\pm ,1}+\sum_{j=1}^{K}\left\vert
p_{j}\right\vert ^{2}A_{\pm ,1}\approx 0
\end{equation}%
and the slowly-varying solutions read%
\begin{equation}
A_{-,1}=A_{-,2}e^{it\delta }~,~A_{+,1}=A_{+,2}e^{-it\delta }\,,
\end{equation}%
where $A_{\pm ,2}$ are some constants and we defined the \emph{frequency shift} due
to the rapidly oscillating terms%
\begin{equation}
\delta =\mathrm{sign}\left( f_{1}\right) \times \frac{\left\vert
f_{1}\right\vert -\sqrt{f_{1}^{2}+4\sum_{j=1}^{K}\left\vert p_{j}\right\vert
^{2}}}{2}  \label{delta}\,.
\end{equation}%
Based on the values presented in Table \ref{table1}, in this work we shall
keep only the terms $\Xi _{-R}$, $\Xi _{-2-R}$ and $\Xi _{-4-R}$ in Eq. (\ref%
{delta}), neglecting all the other (smaller) contributions.

\subsection{Approximate description under resonances}

Now we go back to Eqs. (\ref{e1}) and propose the ansatz%
\begin{equation}
A_{-,1}\left( t\right) =A_{-,2}\left( t\right) e^{it\delta }~,~A_{+,1}\left(
t\right) =A_{+,2}\left( t\right) e^{-it\delta }\,,
\end{equation}%
where $A_{\pm ,2}\left( t\right) $ are slowly varying coefficients (on the typical
timescale $\tau $). As discussed in the previous section, under the
coarse-grained approximation, this ansatz automatically eliminates the
rapidly oscillating part $Q_{f}(t)$ (which merely amounts to introducing the
frequency shift $\delta $). So we are left with the equations%
\begin{equation}
\dot{A}_{-,2}=-iq(t)A_{+,2}~,~\dot{A}_{+,2}=-iq(t)^{\ast }A_{-,2}\,,
\label{sol}
\end{equation}%
where%
\begin{equation}
q(t)\equiv e^{-2i\delta t}Q_{s}(t)=\sum_{k=0}^{\infty }q^{\left( k\right)
}(t)  \label{qw}
\end{equation}%
and $q^{\left( k\right) }$ denotes the $k$-th order slowly-oscillating
terms.

Defining $r\equiv R+2\delta $, the 0th-order terms in $\varepsilon $ are%
\begin{equation}
q^{\left( 0\right) }(t)=\sum_{k=1}^{\infty }{}^{\prime }\Xi _{2k-r}E_{2k-r}\,,
\label{OO}
\end{equation}%
where $\sum_{k=1}^{\infty }{}^{\prime }$ means that \emph{only the terms for which}
$|2\nu k-r|\lesssim |\Xi _{2k-r}|$ \emph{must be retained} [all the remaining
rapidly oscillating terms must be treated as part of $Q_{f}(t)$, and can be
included into the frequency shift $\delta $, Eq. (\ref{delta})]. The initial
terms are%
\begin{equation}
\Xi _{2-r}=e^{i\phi }g_{0}J_{0}^{1}J_{0}^{2}J_{0}^{3}J_{0}^{4}\left(
J_{2}^{0}\tilde{R}_{-}-J_{0}^{0}\tilde{R}_{+}\right)
\end{equation}%
\begin{equation}
\Xi _{4-r}=e^{i\phi }g_{0}J_{0}^{1}J_{0}^{2}J_{0}^{3}J_{0}^{4}\left(
J_{1}^{0}\tilde{R}_{+}-J_{3}^{0}\tilde{R}_{-}\right)
\end{equation}%
\begin{equation}
\Xi _{6-r}=-e^{i\phi }g_{0}\tilde{R}%
_{+}J_{2}^{0}J_{0}^{1}J_{0}^{2}J_{0}^{3}J_{0}^{4}
\end{equation}%
\begin{equation}
\Xi _{8-r}=e^{i\phi }g_{0}\tilde{R}%
_{+}J_{3}^{0}J_{0}^{1}J_{0}^{2}J_{0}^{3}J_{0}^{4}\,.
\end{equation}%
The terms on the right-hand side of Eq. (\ref{OO}) correspond to the
multiphoton atom--field resonances, $2k\nu -r=0$, which occur when the
atomic transition frequency matches the $\left( 2k+1\right) $-photon
resonance:%
\begin{equation}
\Omega _{0}=\nu+2\sqrt{\left( k\nu -\delta \right) ^{2}-g_{0}^{2}}\approx
\left( 2k+1\right) \nu -2\delta -\frac{g_{0}^{2}}{k\nu}\,.
\end{equation}

The first-order terms in $\varepsilon $ are%
\begin{equation}
q^{\left( 1\right) }(t)=\sum_{\eta =\eta _{\Omega },\eta _{g}}\left(
\sum_{k=0}^{\infty }{}^{\prime }\Xi _{-2k-r,\eta }E_{-2k-r+\eta
}+\sum_{k=1}^{\infty }{}^{\prime }\sum_{j=\pm 1}\Xi _{2k-r,j\eta
}E_{2k-r+j\eta }\right) \,,  \label{cof1}
\end{equation}%
where the prime symbol in the sums is a reminder that only the slowly
oscillating terms must be retained; for small values of $k$ the coefficients
are given in the Appendix \ref{appendixA}.

Analogously one can calculate the higher-order terms $q^{\left( k\right)
}(t) $. As an interesting example, below we show the 2-order terms
dependent on the product $\varepsilon _{\Omega }\varepsilon _{g}$ (i. e.,
when both the modulations are imposed simultaneously):%
\begin{eqnarray}
q^{\left( 2\right) }(t) &=&\Xi _{-r,\eta _{\Omega }+\eta _{g}}E_{-r+\eta
_{\Omega }+\eta _{g}}+\Xi _{-r,\eta _{\Omega }-\eta _{g}}E_{-r+\eta _{\Omega
}-\eta _{g}}  \label{cof2} \\
&&+\Xi _{-r,-\eta _{\Omega }+\eta _{g}}E_{-r-\eta _{\Omega }+\eta _{g}}+\Xi
_{-2-r,\eta _{\Omega }+\eta _{g}}E_{-2-r+\eta _{g}+\eta _{\Omega }}
\nonumber \\
&&+\Xi _{-2-r,\eta _{\Omega }-\eta _{g}}E_{-2-r+\eta _{\Omega }-\eta
_{g}}+\Xi _{-2-r,-\eta _{\Omega }+\eta _{g}}E_{-2-r-\eta _{\Omega }+\eta
_{g}}+\cdots,  \nonumber
\end{eqnarray}%
where $\cdots $ denote weaker contributions, and the coefficients of Eq. (\ref%
{cof2}) are given in the Appendix \ref{appendixB}. The right-hand side of Eq. (\ref{cof2}%
) shows that resonances also occur when the combinations $\eta _{\Omega }\pm
\eta _{g}$ match certain values.

Under exact resonance(s), the solution of Eqs. (\ref{sol}) is
\begin{equation}
A_{-,2}=A_{-}\left( 0\right) \cos \left\vert \Xi \right\vert t-ie^{i\xi
}A_{+}\left( 0\right) \sin \left\vert \Xi \right\vert t  \label{f1}
\end{equation}%
\begin{equation}
A_{+,2}=A_{+}\left( 0\right) \cos \left\vert \Xi \right\vert t-ie^{-i\xi
}A_{-}\left( 0\right) \sin \left\vert \Xi \right\vert t~,  \label{f2}
\end{equation}%
where $\Xi \equiv \sum_{k}\Xi _{k}=\left\vert \Xi \right\vert e^{i\xi }$, and
$\Xi _{k}$ stands for the coefficient for which the index of the
exponential $E_{()}$ in Eqs. (\ref{OO}), (\ref{cof1}) and/or (\ref{cof2}) is zero. For
the atom initially in the ground state, the probability of the excited state
reads%
\begin{equation}
P_{e}\left( t\right) =\left\vert \langle e|\psi _{1}(t)\rangle \right\vert
^{2}=\tilde{g}_{0}^{2}\left\vert e^{iT_{2}}e^{-itr}\tilde{R}%
_{+}^{-1/2}A_{+,2}-\tilde{R}_{-}^{-1/2}A_{-,2}\right\vert ^{2}.  \label{pet}
\end{equation}%
Therefore, $\left\vert \Xi \right\vert $ expresses the total transition
rate, which depends on the precise matching of the atomic transition
frequency to a multiphoton resonance condition, as well as adjustment of the modulation
frequencies to the system resonances. Eqs. (\ref{f1}) -- (\ref{f2}) describe
completely the system dynamics at the exact resonances, although in the general case one needs to solve Eqs. (\ref{sol}) numerically. The approximate analytic solution of SRM in the presence of dissipation was given in Refs. \cite{marinho0,marinho1,marinho2}.

\section{Quantum Rabi Model\label{quant}}

The quantum Rabi Hamiltonian \cite{rev,larson} reads%
\begin{equation}
\hat{H}_{Q}=\nu \hat{n}+\Omega \left( t\right) \hat{\sigma}_{e}+\check{g}%
\left( t\right) \left( \hat{a}+\hat{a}^{\dagger }\right) \left( \hat{\sigma}%
_{+}+\hat{\sigma}_{-}\right) =\hat{H}_{0}+2\sum_{k=\Omega ,g}\hat{W}^{\left(
k\right) }\sin \eta _{k}t\,,  \label{HQ}
\end{equation}%
where $\hat{H}_{0}=\nu \hat{n}+\Omega _{0}\hat{\sigma}_{e}+\check{g}%
_{0}\left( \hat{a}+\hat{a}^{\dagger }\right) \left( \hat{\sigma}_{+}+\hat{%
\sigma}_{-}\right) $ is the standard time-independent Rabi Hamiltonian with
bare parameters, and we defined
\begin{equation}
\hat{W}^{\left( \Omega \right) }=\frac{\varepsilon _{\Omega }}{2}\hat{\sigma}%
_{e}~,~\hat{W}^{\left( g\right) }=\frac{\check{\varepsilon}_{g}}{2}\left(
\hat{a}+\hat{a}^{\dagger }\right) \left( \hat{\sigma}_{+}+\hat{\sigma}%
_{-}\right) \,.
\end{equation}%
Here $\hat{a}$ and $\hat{a}^{\dagger }$ are the annihilation and creation
operators of the cavity field, $\hat{n}=\hat{a}^{\dagger }\hat{a}$ is the
photon number operator and $\check{g}\left( t\right) =\check{g}_{0}+\check{%
\varepsilon}_{g}\sin \left( \eta _{g}t\right) $ is the atom--field coupling
parameter in the quantum regime.

Let us briefly see how the quantum Rabi Hamiltonian is related to the semiclassical
one, Eq. (\ref{zer}). In the interaction picture generated by the unitary
operator $\exp \left( -i\nu \hat{n}t\right) $, the Hamiltonian (\ref{HQ})
reads%
\begin{equation}
\hat{H}_{int}=\Omega \left( t\right) \hat{\sigma}_{e}+\check{g}\left(
t\right) \left( \hat{a}e^{-i\nu t}+\hat{a}^{\dagger }e^{i\nu t}\right)
\left( \hat{\sigma}_{+}+\hat{\sigma}_{-}\right) \,.
\end{equation}%
For the cavity field initially in the coherent state $|\alpha \rangle $,
with $\alpha \gg 1$, let us assume that\emph{\ the cavity field is not
altered during the evolution}. Then, the total density operator is $\hat{\rho%
}_{tot}\left( t\right) =\hat{\rho}\left( t\right) \otimes |\alpha \rangle
\langle \alpha |$, where $\hat{\rho}$ is the qubit's density operator, and
the qubit dynamics is governed by the Hamiltonian%
\begin{equation}
\hat{H}_{q}=\mathrm{Tr}_{f}\left[ \hat{H}_{int}\otimes |\alpha \rangle
\langle \alpha |\right] =\Omega \left( t\right) \hat{\sigma}_{e}+\alpha
\check{g}\left( t\right) \left( e^{-i\nu t}+e^{i\nu t}\right) \left(
\hat{\sigma}_{+}+\hat{\sigma}_{-}\right) \,,
\end{equation}%
where $\mathrm{Tr}_{f}$ denotes the partial trace over the field (without
loss of generality, we assumed that $\alpha $ is real). By subtracting the
classical term $\Omega \left( t\right) /2$, one recovers the semiclassical
Hamiltonian (\ref{zer}) with the identification $g\left( t\right) =\alpha
\check{g}\left( t\right) $, which relates the quantum and semiclassical
atom--field coupling parameters. For a rigorous analysis of the transition
from QRM to SRM see Refs. \cite{irish,shus2}.

\subsection{Dressed-states expansion}

Near the multiphoton resonances, when $\Omega _{0}\approx K\nu $, $\check{g}%
_{0}\ll \nu $ and $K$ is a small odd integer, the
non-normalized eigenstates (\textquotedblleft
dressed-states\textquotedblright ) for $n\gg 1$ are, approximately, symmetric and
anti-symmetric combinations of the states $|g,n\rangle $ and $|e,n-K\rangle $%
:%
\begin{eqnarray*}
|S_{n}\rangle  &\approx &\mu _{n}|g,n\rangle +\sqrt{1-\mu _{n}^{2}}%
|e,n-K\rangle -\varsigma _{n}|e,n-1\rangle +\varpi _{n}|g,n-2\rangle +\cdots
\\
|A_{n}\rangle  &\approx &\sqrt{1-\mu _{n}^{2}}|g,n\rangle -\mu
_{n}|e,n-K\rangle -\varpi _{n}|e,n-1\rangle -\varsigma _{n}|g,n-2\rangle
+\cdots ,
\end{eqnarray*}%
where the coefficients $\mu _{n}$, $\varsigma _{n}$, $\varpi _{n}$ etc can
be found from the numeric diagonalization of $\hat{H}_{0}$. The
corresponding eigenenergies, denoted as $S_{n}$ and $A_{n}$, can also be
found numerically \cite{marinho2}. In particular, in the close vicinity of
the $K$-photon resonance, one has $\mu _{n}\approx 2^{-1/2}$, while $%
\varsigma _{n}$, $\varpi _{n}$, etc are much smaller.

We expand the wavefunction corresponding to the Hamiltonian (\ref{HQ}) as%
\begin{equation}
|\psi \rangle =\sum_{n}\left( c_{n}^{\left( S\right)
}e^{-itS_{n}}|S_{n}\rangle +c_{n}^{\left( A\right)
}e^{-itA_{n}}|A_{n}\rangle \right) \,,
\end{equation}%
where the sum runs over all the Hilbert space, and $c_{n}^{\left( S\right) }$
and $c_{n}^{\left( A\right) }$ are the time-dependent probability amplitudes
of the dressed-states. For the initial state $|\psi \left( 0\right) \rangle
=\sum c_{n}|g,n\rangle $, one has
\begin{equation}
c_{n}^{\left( S\right) }\left( 0\right) \approx c_{n}\mu
_{n}~,~c_{n}^{\left( A\right) }\left( 0\right) \approx c_{n}\sqrt{1-\mu
_{n}^{2}}.  \label{frm}
\end{equation}

From the Schr\"{o}dinger equation, we obtain the differential equations%
\begin{eqnarray}
\dot{c}_{m}^{\left( S\right) } &=&-2i\sum_{k=\Omega ,g}\sin \eta
_{k}t\left\{ W_{S_{m},S_{m}}^{\left( k\right) }c_{m}^{\left( S\right)
}+W_{S_{m},A_{m}}^{\left( k\right) }e^{-it\left( A_{m}-S_{m}\right)
}c_{m}^{\left( A\right) }\right.  \nonumber \\
&&\left. +\sum_{n\neq m}\left[ W_{S_{m},S_{n}}^{\left( k\right)
}e^{-it\left( S_{n}-S_{m}\right) }c_{n}^{\left( S\right)
}+W_{S_{m},A_{n}}^{\left( k\right) }e^{-it\left( A_{n}-S_{m}\right)
}c_{n}^{\left( A\right) }\right] \right\}
\end{eqnarray}%
\begin{eqnarray}
\dot{c}_{m}^{\left( A\right) } &=&-2i\sum_{k=\Omega ,g}\sin \left( \eta
_{k}t\right) \left\{ W_{A_{m},A_{m}}^{\left( k\right) }c_{m}^{\left(
A\right) }+W_{A_{m},S_{m}}^{\left( k\right) }e^{-it\left( S_{m}-A_{m}\right)
}c_{m}^{\left( S\right) }\right.  \nonumber \\
&&\left. +\sum_{n\neq m}\left[ W_{A_{m},S_{n}}^{\left( k\right)
}e^{-it\left( S_{n}-A_{m}\right) }c_{n}^{\left( S\right)
}+W_{A_{m},A_{n}}^{\left( k\right) }e^{-it\left( A_{n}-A_{m}\right)
}c_{n}^{\left( A\right) }\right] \right\} \,,
\end{eqnarray}%
where we defined the matrix elements between the dressed-states $%
|A\rangle $ and $|B\rangle $ as%
\begin{equation}
W_{A,B}^{\left( k\right) }\equiv \langle A|\hat{W}^{\left( k\right)
}|B\rangle ~,~\text{for }k=\Omega ,g\,.
\end{equation}

To understand qualitatively the dynamics generated by these equations, let
us define new probability amplitudes $C_{m}^{\left( S\right) }$ and $%
C_{m}^{\left( A\right) }$:%
\begin{eqnarray}
C_{m}^{\left( S\right) }~ &=&c_{m}^{\left( S\right) }\exp \left[
-2i\sum_{p=\Omega ,g}\frac{\cos \eta _{p}t-1}{\eta _{p}}W_{S_{m},S_{m}}^{%
\left( p\right) }\right] \\
C_{m}^{\left( A\right) }\, &=&c_{m}^{\left( A\right) }\exp \left[
-2i\sum_{p=\Omega ,g}\frac{\cos \eta _{p}t-1}{\eta _{p}}W_{A_{m},A_{m}}^{%
\left( p\right) }\right] .
\end{eqnarray}%
Now, one has to solve the equations%
\begin{eqnarray}
\dot{C}_{m}^{\left( S\right) } &=&-2i\sum_{k=\Omega
,g}\sum\nolimits_{n}^{\prime }\sin \left( \eta _{k}t\right) \left\{
W_{S_{m},S_{n}}^{\left( k\right) }\exp \left[ 2i\Upsilon \left(
S_{n},S_{m}\right) \right] e^{-it\left( S_{n}-S_{m}\right) }C_{n}^{\left(
S\right) }\right.  \nonumber \\
&&\left. +W_{S_{m},A_{n}}^{\left( k\right) }\exp \left[ 2i\Upsilon \left(
A_{n},S_{m}\right) \right] e^{-it\left( A_{n}-S_{m}\right) }C_{n}^{\left(
A\right) }\right\}  \label{z1}
\end{eqnarray}%
\begin{eqnarray}
\dot{C}_{m}^{\left( A\right) } &=&-2i\sum_{k=\Omega
,g}\sum\nolimits_{n}^{\prime }\sin \left( \eta _{k}t\right) \left\{
W_{A_{m},S_{n}}^{\left( k\right) }\exp \left[ 2i\Upsilon \left(
S_{n},A_{m}\right) \right] e^{-it\left( S_{n}-A_{m}\right) }C_{n}^{\left(
S\right) }\right.  \nonumber \\
&&\left. +W_{A_{m},A_{n}}^{\left( k\right) }\exp \left[ 2i\Upsilon \left(
A_{n},A_{m}\right) \right] e^{-it\left( A_{n}-A_{m}\right) }C_{n}^{\left(
A\right) }\right\} \,, \label{z2}
\end{eqnarray}%
where $\sum\nolimits_{n}^{\prime }$ denotes the sum over all the
coefficients apart from the one on the left-hand side of the respective
equation, and we defined%
\begin{equation}
\Upsilon \left( A,B\right) \equiv \sum_{p=\Omega ,g}\frac{\cos \eta _{p}t-1}{%
\eta _{p}}\left( W_{A,A}^{\left( p\right) }-W_{B,B}^{\left( p\right)
}\right) \,.
\end{equation}

If $\left\vert W_{A,A}^{\left( p\right) }\right\vert \ll \eta _{p}$ for all
the relevant states $|A\rangle $ and modulation frequencies $\eta _{p}$, one
can expand the exponentials containing $\Upsilon $ in the power series, and
rewrite Eqs. (\ref{z1}) -- (\ref{z2}) in the generic form%
\begin{equation}
C_{m}=\sum_{n\neq m}\left[ \theta _{m,n,1}e^{i\chi _{m,n,1}t}+\theta
_{m,n,2}e^{i\chi _{m,n,2}t}+\cdots \right] C_{n}~,
\end{equation}%
where $C_{n}$ stand for the probability amplitudes, and $\theta _{m,n,l}$
and $\chi _{m,n,l}$ are some constant coefficients (whose exact form is not
important at this point). As discussed in the previous section, the major
contribution to the dynamics comes from the slowly varying terms, for which
$\left\vert \chi _{m,n,l}\right\vert \lesssim \left\vert \theta
_{m,n,l}\right\vert $. Therefore, in our case, the relevant figure of merit is the ratio between the modulation frequency detuning from the transition frequency and the corresponding transition rate. Hence, for the
first-order effects, we define%
\begin{equation}
f_{k}\left( A,B\right) =\left\vert \frac{\left\vert A-B\right\vert -\eta _{k}%
}{W_{A,B}^{\left( k\right) }}\right\vert \,, \label{fm}
\end{equation}%
where $|A\rangle $ and $|B\rangle $ stand for the states within the set $%
\{|S_{n}\rangle ,|A_{m}\rangle \}$, and $A$ and $B$ are the corresponding
eigenenergies. In broad terms, the modulation frequency $\eta
_{k}=\left\vert A-B\right\vert $ causes the transition between the states $%
\{|A\rangle ,|B\rangle \}$ whenever $f_{k}\left( A,B\right) \lesssim 1$. For
higher-order effects, one only needs to replace $W_{A,B}^{\left( k\right) }$
by the appropriate coefficient in Eq. (\ref{fm}).

\begin{figure}[tbh!]
\begin{center}
\includegraphics[width=0.45\textwidth]{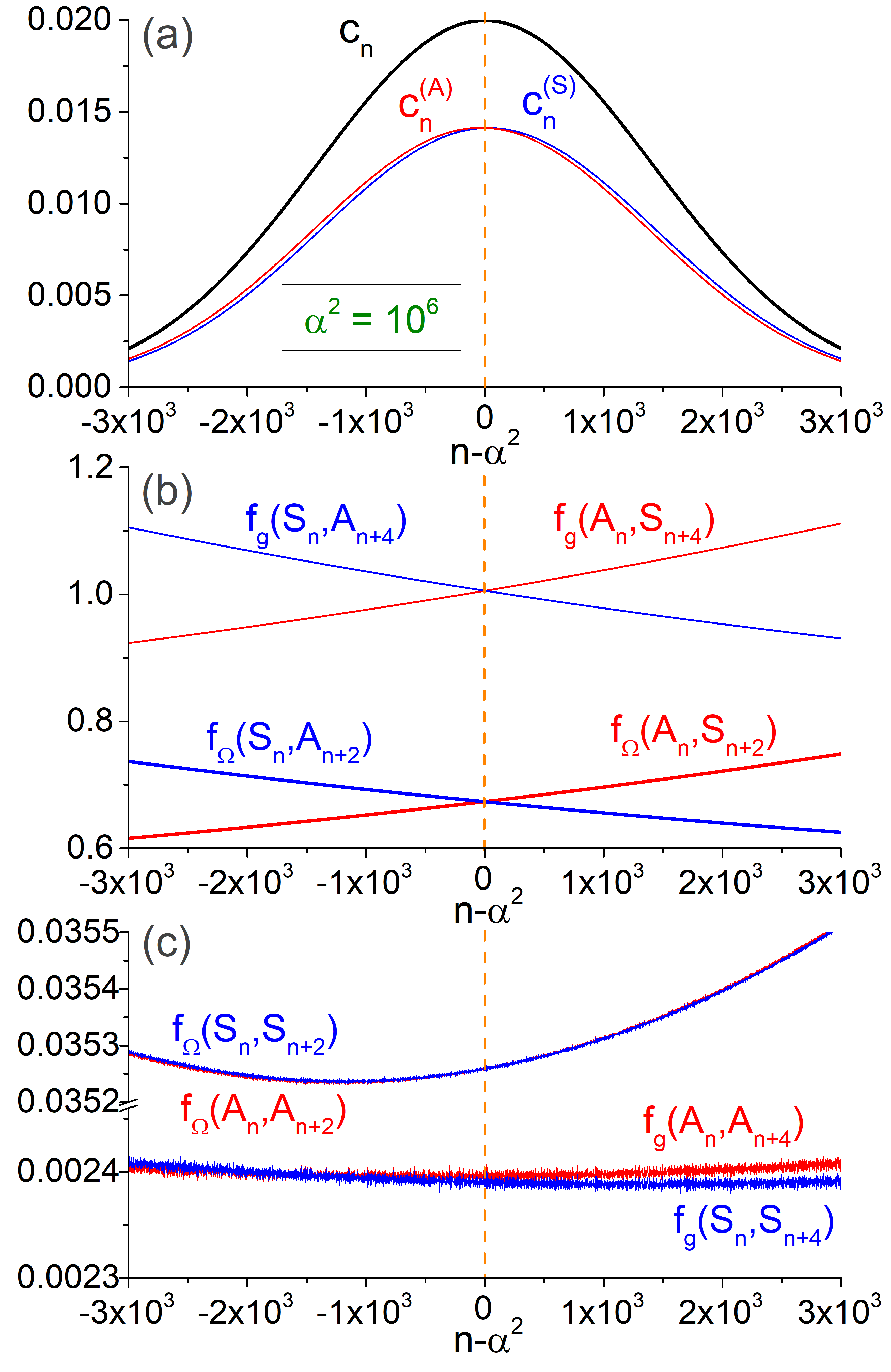} {}
\end{center}
\caption{a) Initial values of the probability amplitudes $c_{n}$ (black
line), $c_{n}^{\left( S\right) }$ (blue line) and $c_{n}^{\left( A\right) }$
(red line) for the initial state $|g,\protect\alpha \rangle $ with $\protect%
\alpha =10^{3}$. b-c) Figures of merit, given by Eq. (\protect\ref{fm}), for
the transitions between different dressed-states and parameters: $\check{g}%
_{0}=0.1\protect\nu /\protect\alpha $, $\Omega _{0}=2.984973\protect\nu $, $%
\protect\varepsilon _{\Omega }=0.02\Omega _{0}$, $\check{\protect\varepsilon}%
_{g}=-0.02\check{g}_{0}$, $\protect\eta _{\Omega }=2\protect\nu $, $%
\protect\eta _{g}=4\protect\nu $. Smaller values of $f_{k}\left(
A,B\right) $ indicate stronger effective coupling between the states $%
|A\rangle $ and $|B\rangle .$}
\label{fig1}
\end{figure}

In Fig. \ref{fig1} we consider the initial state $|\psi \left(
0\right) \rangle =|g\rangle \otimes |\alpha \rangle $, with the atom in the
ground state and the cavity in the coherent state, where $\alpha =10^{3}$
and the parameters of the Hamiltonian $\hat{H}_{Q}$ are $\check{g}%
_{0}=0.1\nu /\alpha $, $\Omega _{0}=2.984973\nu $, $\varepsilon _{\Omega
}=0.02\Omega _{0}$, $\check{\varepsilon}_{g}=-0.02\check{g}_{0}$, $\eta
_{\Omega }=2\nu $ and $\eta _{g}=4\nu $. Fig. \ref{fig1}a shows the
initial values of $c_{n}=e^{-\alpha ^{2}/2}\alpha ^{n}/\sqrt{n!}$ (black
line), together with $c_{n}^{\left( S\right) }$ (blue line) and $%
c_{n}^{\left( A\right) }$ (red line) obtained from Eq. (\ref{frm}). We see
that for $n>\alpha ^{2}$ we have $c_{n}^{\left( S\right) }>c_{n}^{\left(
A\right) }$, while for $n<\alpha ^{2}$ we have $c_{n}^{\left( A\right)
}>c_{n}^{\left( S\right) }$. In Fig. \ref{fig1}b we plot $f_{\Omega }\left(
S_{n},A_{n+2}\right) $, $f_{\Omega }\left( A_{n},S_{n+2}\right) $, $%
f_{g}\left( S_{n},A_{n+4}\right) $ and $f_{g}\left( A_{n},S_{n+4}\right) $;
in Fig. \ref{fig1}c we plot $f_{\Omega }\left( S_{n},S_{n+2}\right) $, $%
f_{\Omega }\left( A_{n},A_{n+2}\right) $, $f_{g}\left( S_{n},S_{n+4}\right) $
and $f_{g}\left( A_{n},A_{n+4}\right) $. We see that for $n>\alpha ^{2}$ one
has $f_{\Omega }\left( S_{n},A_{n+2}\right) <f_{\Omega }\left(
A_{n},S_{n+2}\right) $ and $f_{g}\left( S_{n},A_{n+4}\right) <\left(
A_{n},S_{n+4}\right) $, meaning that for such values of $n$ the transitions $%
|S_{n}\rangle \rightarrow |A_{n+2}\rangle $ and $|S_{n}\rangle \rightarrow
|A_{n+4}\rangle $ are more favorable than $|A_{n}\rangle \rightarrow
|S_{n+2}\rangle $ and $|A_{n}\rangle \rightarrow |S_{n+4}\rangle $. And
precisely for such values of $n$ one has $c_{n}^{\left( S\right)
}>c_{n+2}^{\left( A\right) },c_{n+4}^{\left( A\right) }$. For $n<\alpha ^{2}$
one has the opposite scenario. Therefore, the external modulations with
frequencies $\eta _{\Omega }=2\nu $ and $\eta _{g}=4\nu $ induce the
transitions that increase the number of photons, compared to the
unperturbed case. Since $f_{\Omega }\left( S_{n},S_{n+2}\right) \approx
f_{\Omega }\left( A_{n},A_{n+2}\right) $ and $f_{g}\left(
S_{n},S_{n+4}\right) \approx f_{g}\left( A_{n},A_{n+4}\right) $, while $%
c_{n}^{\left( S,A\right) }\approx c_{n+2}^{\left( S,A\right) }\approx
c_{n+4}^{\left( S,A\right) }$ (as seen in Figs. \ref{fig1}a and \ref{fig1}%
c), the transitions between the states $|A_{n}\rangle \rightarrow
|A_{n+2}\rangle $, $|S_{n}\rangle \rightarrow |S_{n+2}\rangle $, $%
|A_{n}\rangle \rightarrow |A_{n+4}\rangle $ and $|S_{n}\rangle \rightarrow
|S_{n+4}\rangle $ do not contribute significantly to the dynamics in this
case. Hence, Fig. \ref{fig1} explains qualitatively how the resonant
external modulation can alter the system dynamics for the initial coherent
state with a large amplitude. For the initial vacuum state, the dynamics of
photon generation due to parametric modulation has been studied in numerous papers \cite%
{j1,j3,j2,tom,messina,j4}.

\subsection{Approximate equations for 1- and 2-order effects}

Now, let us see the approximate equations that describe the system dynamics
during the first and second-order resonances. For the sake of illustrations,
first we assume that $\eta _{\Omega }=2\nu $ and $\eta _{g}=0$. Neglecting
the rapidly-oscillating terms, for which $f_{\Omega }\left( A,B\right) \gg 1$%
, to the first order in $\varepsilon _{\Omega }$ we obtain equations of the
form%
\begin{eqnarray}
\dot{C}_{m}^{\left( S\right) } &\simeq &W_{S_{m},S_{m-2}}^{\left( \Omega
\right) }e^{it\left( \left\vert S_{m}-S_{m-2}\right\vert -\eta _{\Omega
}\right) }C_{m-2}^{\left( S\right) }-W_{S_{m},S_{m+2}}^{\left( \Omega
\right) }e^{-it\left( \left\vert S_{m+2}-S_{m}\right\vert -\eta _{\Omega
}\right) }C_{m+2}^{\left( S\right) }  \label{z3} \\
&&+W_{S_{m},A_{m-2}}^{\left( \Omega \right) }e^{it\left( \left\vert
S_{m}-A_{m-2}\right\vert -\eta _{\Omega }\right) }C_{m-2}^{\left( A\right)
}-W_{S_{m},A_{m+2}}^{\left( \Omega \right) }e^{-it\left( \left\vert
A_{m+2}-S_{m}\right\vert -\eta _{\Omega }\right) }C_{m+2}^{\left( A\right) }
\nonumber
\end{eqnarray}%
\begin{eqnarray}
\dot{C}_{A,m} &\simeq &W_{A_{m},S_{m-2}}^{\left( \Omega \right) }e^{it\left(
\left\vert A_{m}-S_{m-2}\right\vert -\eta _{\Omega }\right) }C_{m-2}^{\left(
S\right) }-W_{A_{m},S_{m+2}}^{\left( \Omega \right) }e^{-it\left( \left\vert
S_{m+2}-A_{m}\right\vert -\eta _{\Omega }\right) }C_{m+2}^{\left( S\right) }
\label{z4} \\
&&+W_{A_{m},A_{m-2}}^{\left( \Omega \right) }e^{it\left( \left\vert
A_{m}-A_{m-2}\right\vert -\eta _{\Omega }\right) }C_{m-2}^{\left( A\right)
}-W_{A_{m},A_{m+2}}^{\left( \Omega \right) }e^{-it\left( \left\vert
A_{m+2}-A_{m}\right\vert -\eta _{\Omega }\right) }C_{m+2}^{\left( A\right)
}\,.  \nonumber
\end{eqnarray}%
If $\eta _{g}\neq 0$, one should simply add the terms proportional to $%
W_{A,B}^{\left( g\right) }$ on the RHS of Eqs. (\ref{z3}) -- (\ref{z4}).
For example, the additional modulation with $\eta _{g}=4\nu $ contributes
with terms proportional to $C_{m\pm 4}^{\left( S\right) }$ and $C_{m\pm
4}^{\left( A\right) }$ on the RHS of Eqs. (\ref{z3}) -- (\ref{z4}), which
correspond to the coupling between the states $\{|S_{m}\rangle
,|A_{m}\rangle \}\ $and $\{|S_{m\pm 4}\rangle ,|A_{m\pm 4}\rangle \}$.

For illustrating the second-order effects, let us consider the coupling
among the states $\left\{ |S_{m}\rangle ,|A_{m}\rangle ,|S_{m\pm 2}\rangle
,|A_{m\pm 2}\rangle \right\} $ for $\eta _{\Omega }+\eta _{g}=2\nu $. We
obtain%
\begin{eqnarray*}
i\dot{C}_{m}^{\left( S\right) } &\simeq &\left[ \frac{W_{S_{m},S_{m+2}}^{%
\left( \Omega \right) }}{\eta _{g}}\left( W_{S_{m+2},S_{m+2}}^{\left(
g\right) }-W_{S_{m},S_{m}}^{\left( g\right) }\right) +\frac{%
W_{S_{m},S_{m+2}}^{\left( g\right) }}{\eta _{\Omega }}\left(
W_{S_{m+2},S_{m+2}}^{\left( \Omega \right) }-W_{S_{m},S_{m}}^{\left( \Omega
\right) }\right) \right]  \\
&&\times e^{-it\left( \left\vert S_{m+2}-S_{m}\right\vert -\eta _{g}-\eta
_{\Omega }\right) }C_{m+2}^{\left( S\right) } \\
&&+\left[ \frac{W_{S_{m},A_{m+2}}^{\left( \Omega \right) }}{\eta _{g}}\left(
W_{A_{m+2},A_{m+2}}^{\left( g\right) }-W_{S_{m},S_{m}}^{\left( g\right)
}\right) +\frac{W_{S_{m},A_{m+2}}^{\left( g\right) }}{\eta _{\Omega }}\left(
W_{A_{m+2},A_{m+2}}^{\left( \Omega \right) }-W_{S_{m},S_{m}}^{\left( \Omega
\right) }\right) \right]  \\
&&\times e^{-it\left( \left\vert A_{m+2}-S_{m}\right\vert -\eta _{g}-\eta
_{\Omega }\right) }C_{m+2}^{\left( A\right) } \\
&&+\left[ \frac{W_{S_{m},S_{m-2}}^{\left( \Omega \right) }}{\eta _{g}}\left(
W_{S_{m-2},S_{m-2}}^{\left( g\right) }-W_{S_{m},S_{m}}^{\left( g\right)
}\right) +\frac{W_{S_{m},S_{m-2}}^{\left( g\right) }}{\eta _{\Omega }}\left(
W_{S_{m-2},S_{m-2}}^{\left( \Omega \right) }-W_{S_{m},S_{m}}^{\left( \Omega
\right) }\right) \right]  \\
&&\times e^{it\left( \left\vert S_{m}-S_{m-2}\right\vert -\eta _{g}-\eta
_{\Omega }\right) }C_{m-2}^{\left( S\right) } \\
&&+\left[ \frac{W_{S_{m},A_{m-2}}^{\left( \Omega \right) }}{\eta _{g}}\left(
W_{A_{m-2},A_{m-2}}^{\left( g\right) }-W_{S_{m},S_{m}}^{\left( g\right)
}\right) +\frac{W_{S_{m},A_{m-2}}^{\left( g\right) }}{\eta _{\Omega }}\left(
W_{A_{m-2},A_{m-2}}^{\left( \Omega \right) }-W_{S_{m},S_{m}}^{\left( \Omega
\right) }\right) \right]  \\
&&\times e^{it\left( \left\vert S_{m}-A_{m-2}\right\vert -\eta _{g}-\eta
_{\Omega }\right) }C_{m-2}^{\left( A\right) }
\end{eqnarray*}%
\begin{eqnarray*}
i\dot{C}_{m}^{\left( A\right) } &\simeq &\left[ \frac{W_{A_{m},S_{m+2}}^{%
\left( \Omega \right) }}{\eta _{g}}\left( W_{S_{m+2},S_{m+2}}^{\left(
g\right) }-W_{A_{m},A_{m}}^{\left( g\right) }\right) +\frac{%
W_{A_{m},S_{m+2}}^{\left( g\right) }}{\eta _{\Omega }}\left(
W_{S_{m+2},S_{m+2}}^{\left( \Omega \right) }-W_{A_{m},A_{m}}^{\left( \Omega
\right) }\right) \right]  \\
&&\times e^{-it\left( \left\vert S_{m+2}-A_{m}\right\vert -\eta _{g}-\eta
_{\Omega }\right) }C_{n}^{\left( S\right) } \\
&&+\left[ \frac{W_{A_{m},A_{m+2}}^{\left( \Omega \right) }}{\eta _{g}}\left(
W_{A_{m+2},A_{m+2}}^{\left( g\right) }-W_{A_{m},A_{m}}^{\left( g\right)
}\right) +\frac{W_{A_{m},A_{m+2}}^{\left( g\right) }}{\eta _{\Omega }}\left(
W_{A_{m+2},A_{m+2}}^{\left( \Omega \right) }-W_{A_{m},A_{m}}^{\left( \Omega
\right) }\right) \right]  \\
&&\times e^{-it\left( \left\vert A_{m+2}-A_{m}\right\vert -\eta _{g}-\eta
_{\Omega }\right) }C_{n}^{\left( A\right) } \\
&&+\left[ \frac{W_{A_{m},S_{m-2}}^{\left( \Omega \right) }}{\eta _{g}}\left(
W_{S_{m-2},S_{m-2}}^{\left( g\right) }-W_{A_{m},A_{m}}^{\left( g\right)
}\right) +\frac{W_{A_{m},S_{m-2}}^{\left( g\right) }}{\eta _{\Omega }}\left(
W_{S_{m-2},S_{m-2}}^{\left( \Omega \right) }-W_{A_{m},A_{m}}^{\left( \Omega
\right) }\right) \right]  \\
&&\times e^{it\left( \left\vert A_{m}-S_{m-2}\right\vert -\eta _{g}-\eta
_{\Omega }\right) }C_{n}^{\left( S\right) } \\
&&+\left[ \frac{W_{A_{m},A_{m-2}}^{\left( \Omega \right) }}{\eta _{g}}\left(
W_{A_{m-2},A_{m-2}}^{\left( g\right) }-W_{A_{m},A_{m}}^{\left( g\right)
}\right) +\frac{W_{A_{m},A_{m-2}}^{\left( g\right) }}{\eta _{\Omega }}\left(
W_{A_{m-2},A_{m-2}}^{\left( \Omega \right) }-W_{A_{m},A_{m}}^{\left( \Omega
\right) }\right) \right]  \\
&&\times e^{it\left( \left\vert A_{m}-A_{m-2}\right\vert -\eta _{g}-\eta
_{\Omega }\right) }C_{n}^{\left( A\right) }
\end{eqnarray*}%
As expected, the transition rates between the states depend on the product $%
\varepsilon _{\Omega }\varepsilon _{g}$, so under the realistic scenario of
modulations with small amplitudes, the second-order effects are
significantly weaker than the first-order ones \cite{tom}.

\section{Numeric results\label{numeric}}

In this section we analyze the system dynamics for the initial state $%
|g\rangle \otimes |\alpha \rangle $ and two values of the initial
average photon number: $\alpha ^{2}=5\times 10^{3}$ and $\alpha ^{2}=3\times
10^{4}$. To compare the predictions of the QRM in the semiclassical
regime (i.e., for large values of $\alpha $) to the ones of SRM, we solved numerically the Schr%
\"{o}dinger equation corresponding to Eq. (\ref{HQ}) using the
Runge-Kutta-Verner fifth-order and sixth-order method. We assumed the truncated
subspace
\begin{equation}
|\psi (t)\rangle =\sum_{n=N_{\min }}^{N_{\max }}\left( A_{n}|g,n\rangle
+B_{n}|e,n\rangle \right) ~,\label{fra}
\end{equation}%
where $N_{\min }$ and $N_{\max }$ are such that for the initial coherent
state one has $\langle n|\alpha \rangle <10^{-18}$ for $n\notin \left[
N_{\min },N_{\max }\right] $. Since we treat a dissipationless case with an
exchange of a small number of excitations between the atom and the field,
Eq. (\ref{fra}) a fair approximation for our problem.

\subsection{3-photon qubit excitation without modulation}

\begin{figure}[tbh!]
\begin{center}
\includegraphics[width=0.65\textwidth]{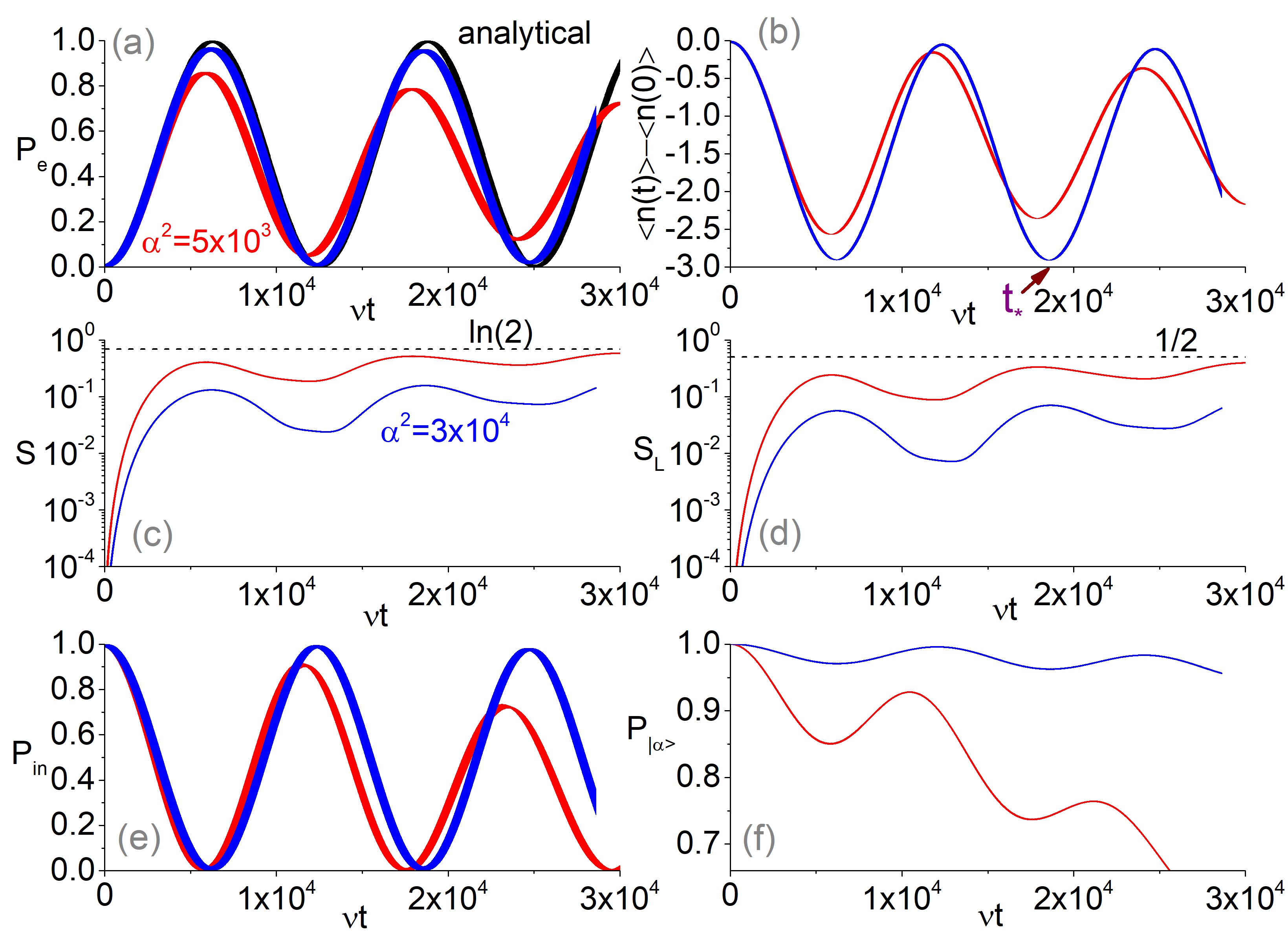} {}
\end{center}
\caption{Atom--field dynamics at the three-photon resonance without external
modulation. Parameters: $\check{g}_{0}=0.1\protect\nu /\protect\alpha $, $%
\Omega _{0}=2.98497\protect\nu $, $\protect\alpha ^{2}=5\times 10^{3}$ (red
lines) and $\protect\alpha ^{2}=3\times 10^{4}$ (blue lines). a) Qubit
excitation probability. Black lines depicts the analytic solution, Eq. (%
\protect\ref{pet}). b) Variation of the average photon number. $t_{\ast
}=18910\protect\nu ^{-1}$ is the instant of time for which the photon number
statistics is analyzed in Fig. \protect\ref{fig3}. c) Von Neumann entropy of
either subsystem. d) Linear entropy of either subsystem. e) Probability of
the initial state, $|\langle \protect\psi (t)|g,\protect\alpha (t)\rangle
|^{2}$, where $\protect\alpha (t)=\protect\alpha e^{-i\protect\nu t}$. f)
Probability of the coherent state $|\protect\alpha (t)\rangle $.}
\label{fig2}
\end{figure}

In Fig. \ref{fig2} we consider the three-photon atomic excitation in the
absence of any external modulation, considering parameters $g_{0}=0.1\nu $, $%
\check{g}(t)=g(t)/\alpha $ and $\Omega _{0}=2.98497\nu $, which correspond
to the exact 3-photon resonance in SRM (the comparison between SRM and QRM for the standard
one-photon resonance has been described in \cite{luo,acosta1}). To
illustrate the transition from the quantum to semiclassical behavior, we
considered two values of the initial average photon number: $\alpha
^{2}=5\times 10^{3}$ (red lines) and $\alpha ^{2}=3\times 10^{4}$ (blue
lines). Panel \ref{fig2}a shows the behavior of the atomic excitation
probability, $P_{e}(t)=\limfunc{Tr}[|e\rangle \langle e|\hat{\rho}]$, where $%
\hat{\rho}=|\psi (t)\rangle \langle \psi (t)|$ is the density matrix of the
total system. The black line depicts the analytic formula (\ref{pet}) for
the SRM. We verified that in all cases studied in this paper the formula (%
\ref{pet}) is indistinguishable from the numeric solution of the Hamiltonian
(\ref{zer}) within the thickness of the lines, so the numeric solution of
SRM is not shown. We see that for $\alpha ^{2}=5\times 10^{3}$ the classical
and quantum behaviors are significantly different, while for $\alpha
^{2}=3\times 10^{4}$ the behavior is almost identical, at least for
initial times. For long times, inevitably, the quantum and classical
behaviors differ due to the collapse-revival phenomenon \cite{scully,eberly}, which occurs faster
for smaller values of $\alpha $ and is not captured by SRM.

In the remaining five panels of Fig. \ref{fig2} we study the quantities that
only exist in QRM. In the panel \ref{fig2}b we plot the variation of the
average photon number, $\left\langle n(t)\right\rangle -\left\langle
n(0)\right\rangle $, where $\left\langle n\right\rangle =\limfunc{Tr}(\hat{n}%
\hat{\rho})$. This plot illustrates that the atomic excitation is
accompanied by the decrease of the average number of photons in the cavity:
for $\alpha ^{2}=3\times 10^{4}$ it decreases by almost three photons, while
for $\alpha ^{2}=5\times 10^{3}$ the variation is smaller (of the order of $%
-2.5$) because not all the Fock states undergo transitions. To study the
atom--field entanglement, in panels \ref{fig2}c and \ref{fig2}d we present
the behavior of the von Neumann entropy, $S=-\left\langle \ln \hat{\rho}%
_{sub}\right\rangle $, and the linear entropy, $S_{L}=1-\limfunc{Tr}(\hat{%
\rho}_{sub}^{2})$ [trivially related to the purity $\limfunc{Tr}(\hat{\rho}%
_{sub}^{2})$]. $\hat{\rho}_{sub}$ is the reduced density operator of either
of the subsystems, since from the Schmidt decomposition for pure states,
they have the same eigenvalues. Nonzero values of $S$ or $S_{L}$ attest the
atom--field entanglement. As expected, the excitation of the atom leads to
the atom--field entanglement, and for more intense (more \textquotedblleft
classical\textquotedblright ) coherent states, the degree of entanglement is
smaller for a given instant of time. Furthermore, when the atom returns to
the ground state due to the reemission of photons, the entropies only
exhibit local minima, but do not vanish. Indeed, the entropies increase over
time, undergoing oscillations, attaining the maximum allowed values $%
S=\ln 2$ and $S_{L}=1/2$ for large times (for the time interval considered in
Fig. \ref{fig2}, this can only be seen for $\alpha ^{2}=5\times 10^{3}$),
indicating the formation of maximally entangled states. We note that our
approach neglects the damping and dephasing mechanisms, so the numeric
results for large times, $\gamma _{\max }t\gtrsim 1$, are not meaningful
from the experimental point of view (we use the symbol $\gamma _{\max }$ to
denote the largest of the damping and dephasing rates).

\begin{figure}[tbh!]
\begin{center}
\includegraphics[width=0.99\textwidth]{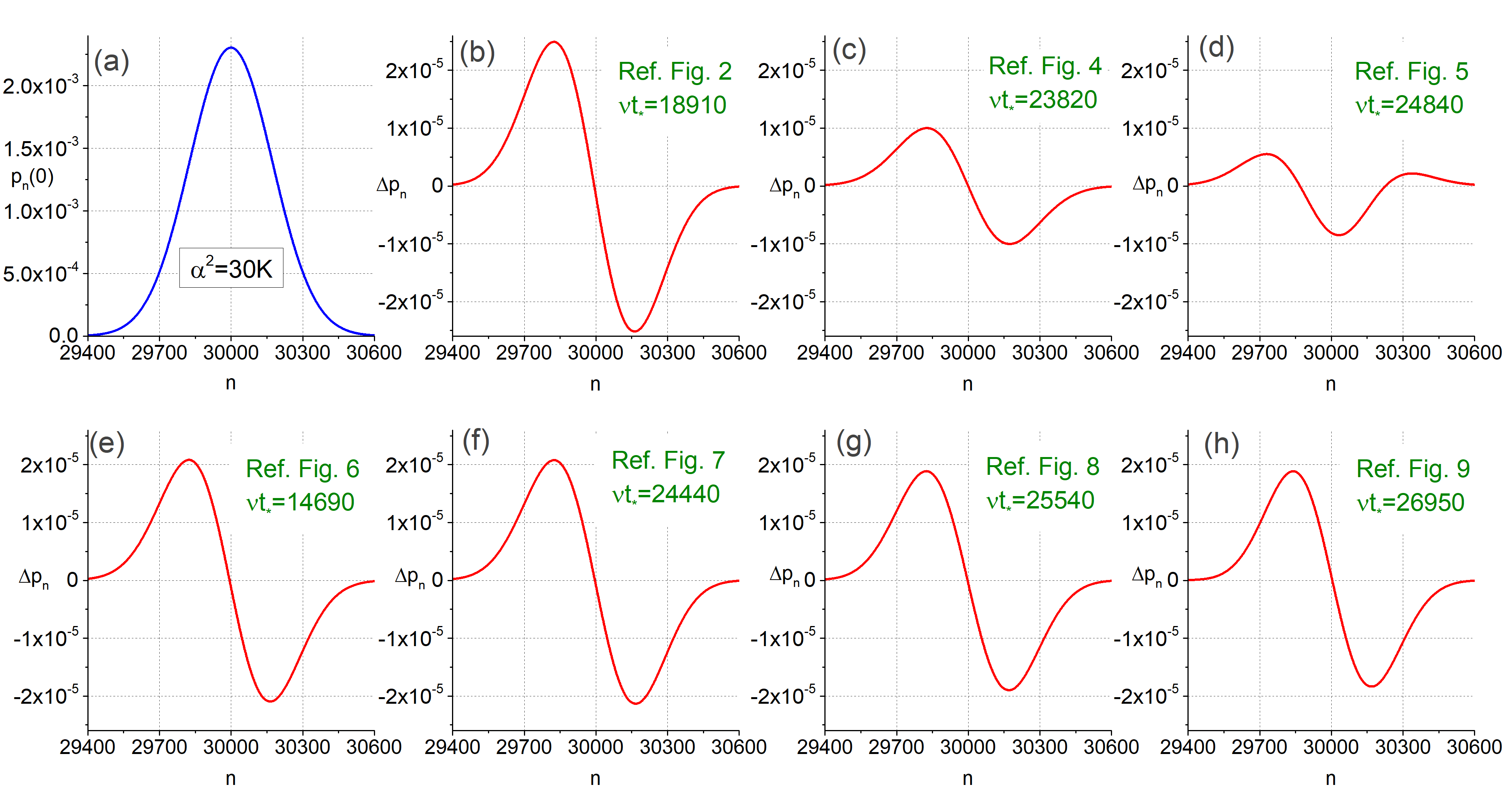} {}
\end{center}
\caption{a) Photon number probability distribution for the initial coherent
state $|\protect\alpha \rangle $, where $\protect\alpha ^{2}=3\times 10^{4}$%
. b-h) Difference between the photon number probability distributions at the
instant of time $t_{\ast }$ and the initial time, corresponding to the
parameters of Figs. \protect\ref{fig2} and \protect\ref{fig4} -- \protect
\ref{fig9}.}
\label{fig3}
\end{figure}

To better understand how the atom alters cavity field state, in the panels %
\ref{fig2}e and \ref{fig2}f we plot the probability of the system state
remaining in the initial state, $P_{in}=\left\vert \langle \psi
(t)|g,\alpha \left( t\right) \rangle \right\vert ^{2}$, and the probability
of the cavity field remaining in the original coherent state $P_{|\alpha \rangle }=%
\limfunc{Tr}(|\alpha \left( t\right) \rangle \langle \alpha \left( t\right) |%
\hat{\rho})$, where we took into account the free evolution of the cavity
field, defining $\alpha \left( t\right) =\alpha e^{-i\nu t}$ (in the absence
of the atom, this would be the solution of the Schr\"{o}dinger equation).
From the panel \ref{fig2}e we see that at the moments of the maximum
excitation of the qubit, the probability $P_{in}$ goes to zero, meaning that
the system state is indeed driven away from the initial state. Moreover, $%
P_{|\alpha \rangle }$ decreases over time, exhibiting oscillations with the
periodicity of the qubit oscillations. As expected, the decay of $P_{|\alpha
\rangle }$ is much slower for larger values of $\alpha $, meaning that for
sufficiently intense coherent states the assumption that the field remains
in the original coherent state becomes partially justified for initial times.

In Fig. \ref{fig3} we illustrate how the photon number distribution is
modified due to the interaction with the atom. The panel \ref{fig3}a shows
the photon number distribution of the original field state, $p_{n}\left(
0\right) =e^{-\alpha ^{2}}\alpha ^{2n}/n!$, for $\alpha ^{2}=3\times 10^{4}$%
. The panel \ref{fig3}a shows the modification of the photon number
probability distribution, $\Delta p_{n}=p_{n}\left( t_{\ast }\right) -p_{n}\left(
0\right) $, for the time instant $\nu t_{\ast }=18910$ (this instant of time
is indicated by a small arrow in Fig. \ref{fig2}b), where $p_{n}\left(
t\right) =\limfunc{Tr}\left[ |n\rangle \langle n|\hat{\rho}\left( t\right) %
\right] $. We see that the atomic excitation is accompanied by a slight
decrease of the photon number probability for $n>\alpha ^{2}$ and a
corresponding increase for $n<\alpha ^{2}$, which explains the decrease of
the average number of photons.

\subsection{Modulation: first-order effects}

\begin{figure}[tbh!]
\begin{center}
\includegraphics[width=0.65\textwidth]{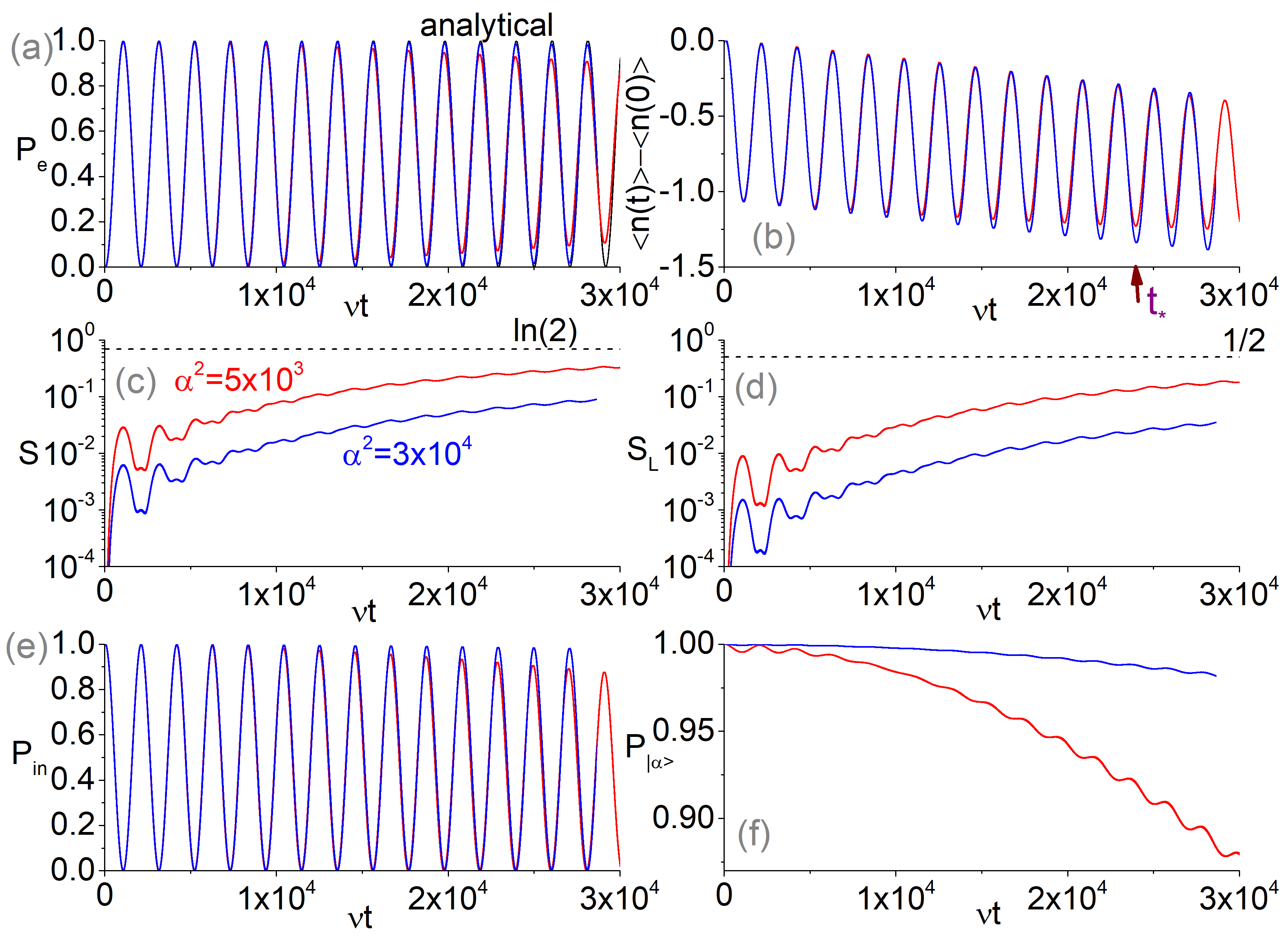} {}
\end{center}
\caption{Similar to Fig. \protect\ref{fig2}, but in the presence of $\Omega $%
-modulation with parameters $\protect\varepsilon _{\Omega }=0.02\Omega _{0}$
and $\protect\eta _{\Omega }=2\protect\nu $. Here $t_{\ast }=23820\protect%
\nu ^{-1}$.}
\label{fig4}
\end{figure}
\begin{figure}[tbh!]
\begin{center}
\includegraphics[width=0.65\textwidth]{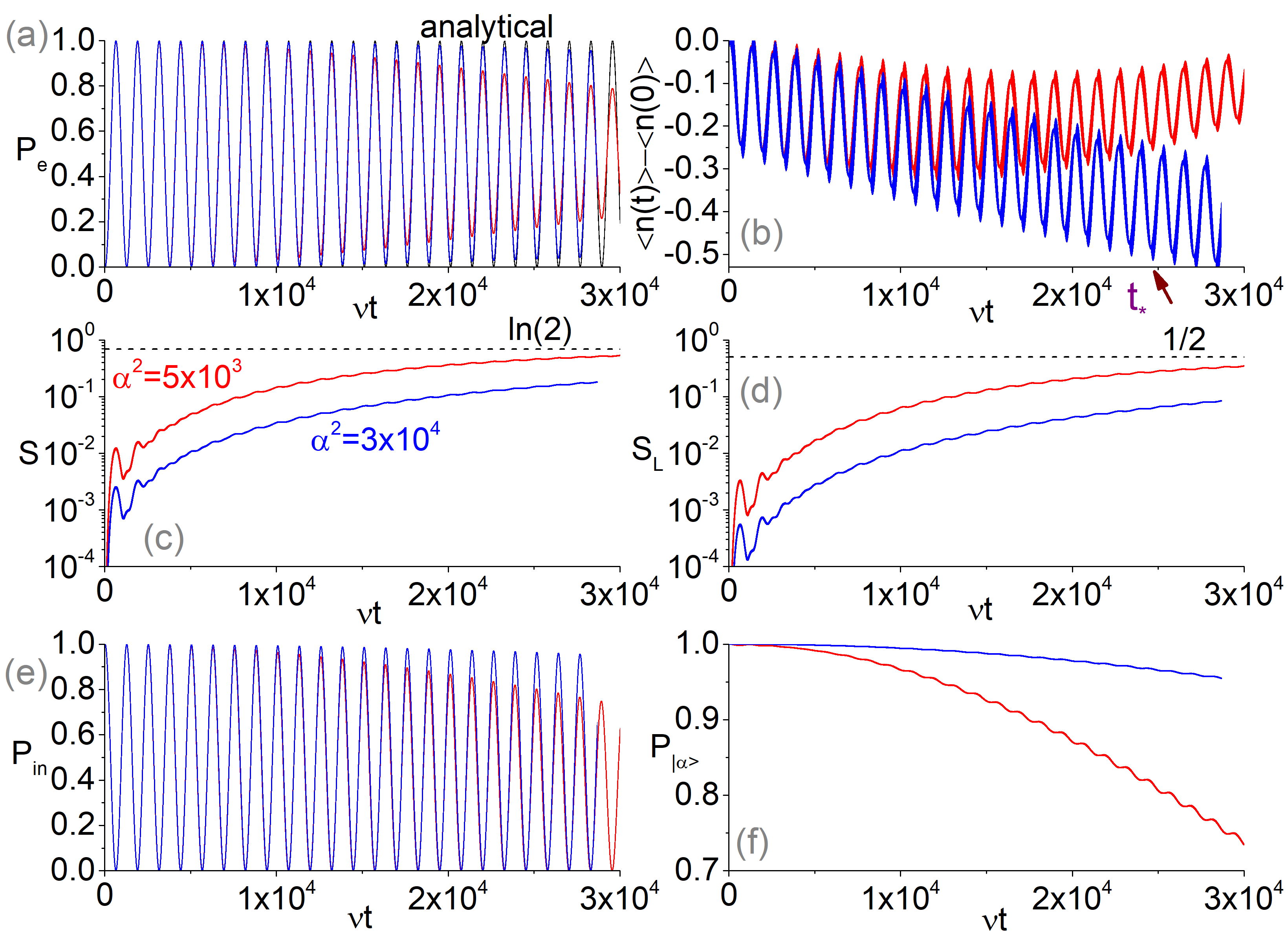} {}
\end{center}
\caption{Similar to Fig. \protect\ref{fig4}, but with the addition of the simultaneous $g$%
-modulation. Parameters: $%
\check{\protect\varepsilon}_{g}=-0.02\check{g}_{0}$, $\protect\eta _{g}=4%
\protect\nu $ and $t_{\ast }=24840\protect\nu ^{-1}$.}
\label{fig5}
\end{figure}

In Fig. \ref{fig4} we consider the parameters of Fig. \ref{fig2}
supplemented with the $\Omega $-modulation with parameters $\varepsilon
_{\Omega }=0.02\Omega _{0}$ and $\eta _{\Omega }=2\nu $. As expected, the
period of the atomic Rabi oscillations decreases, and the dynamics is well
described by QRM already for $\alpha ^{2}=5\times 10^{3}$. But now the
photon-number behavior is different, because the external modulation
activates the additional transitions $|\varphi _{n}^{(1,2)}\rangle
\rightarrow |\varphi _{n+2}^{(1,2)}\rangle $ (where $\varphi $ stands for $S$
and $A$), as seen from Fig. \ref{fig4}b.

In Fig. \ref{fig5} we add the $g$-modulation with parameters $\varepsilon
_{g}=-0.02g_{0}$ and $\eta _{g}=4\nu $ to the previous scenario. The qubit
oscillations become even faster, and the photon dynamics is different again,
since now the additional transition $|\varphi _{n}^{(1,2)}\rangle
\rightarrow |\varphi _{n+4}^{(1,2)}\rangle $ are activated.

\subsection{Modulation: second-order effects}

\begin{figure}[tbh!]
\begin{center}
\includegraphics[width=0.65\textwidth]{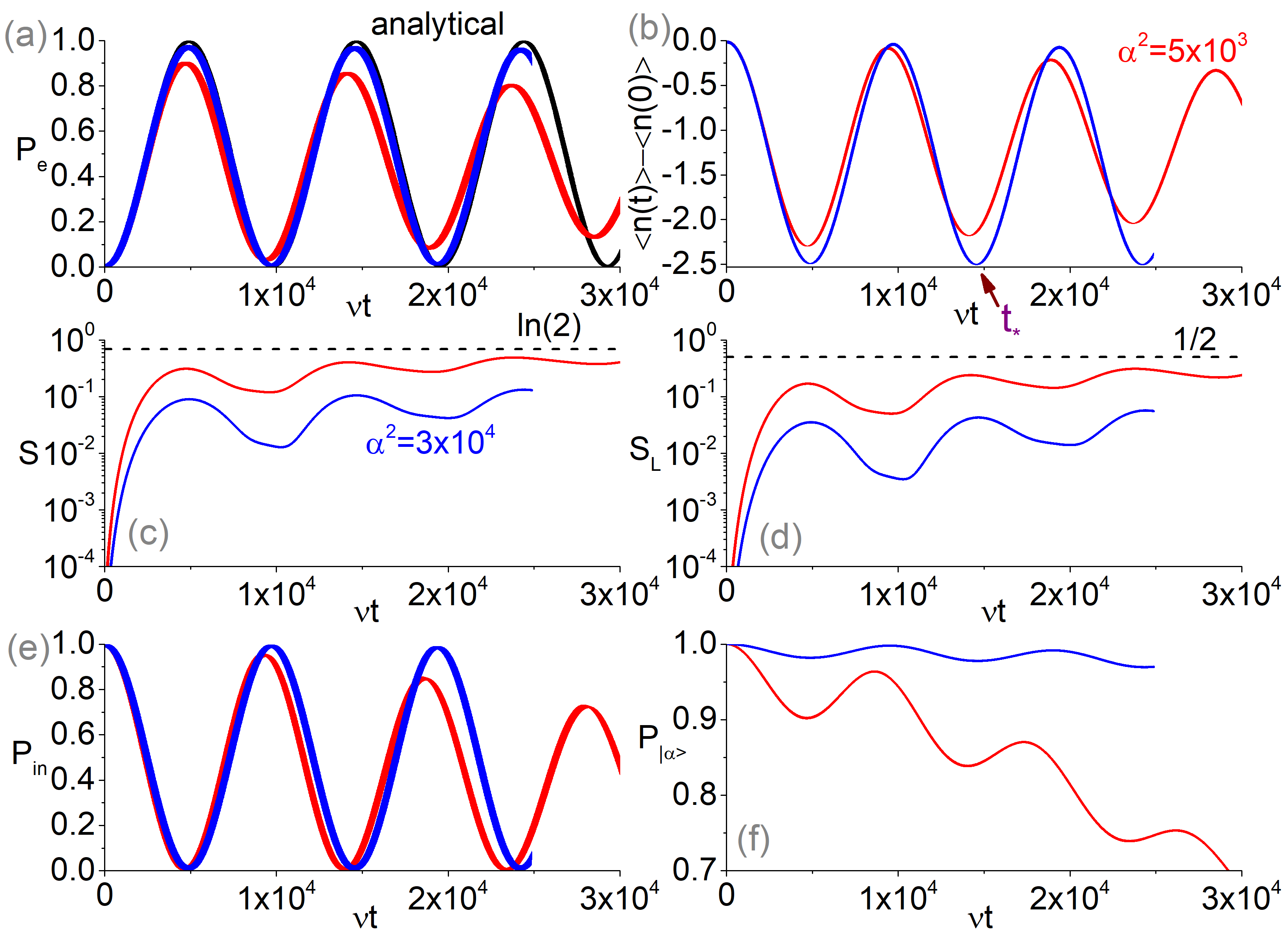} {}
\end{center}
\caption{Similar to Fig. \protect\ref{fig2}, but in the presence of
simultaneous $\Omega $- and $g$-modulations, illustrating the second-order
effects. Parameters: $\protect\varepsilon _{\Omega }=0.02\Omega _{0}$, $%
\protect\varepsilon _{g}=-0.02g_{0}$, $\protect\eta _{\Omega }=0.4\protect%
\nu $ and $\protect\eta _{g}=1.6\protect\nu $, so that $\protect\eta _{g}+%
\protect\eta _{\Omega }=2\protect\nu $. Here $t_{\ast }=14690\protect\nu %
^{-1}$.}
\label{fig6}
\end{figure}
\begin{figure}[tbh!]
\begin{center}
\includegraphics[width=0.65\textwidth]{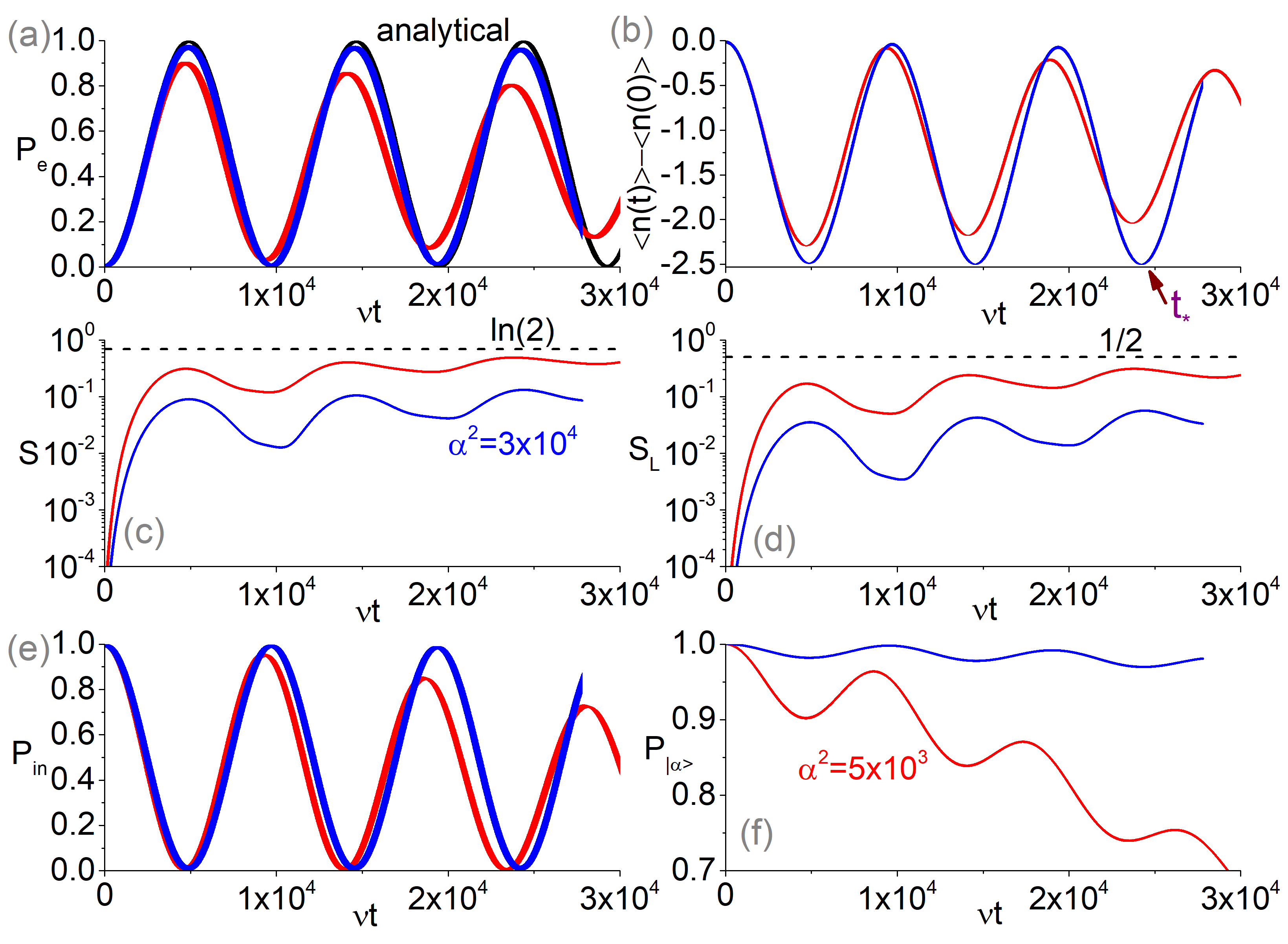} {}
\end{center}
\caption{Similar to Fig. \protect\ref{fig6}, but for $\protect\eta _{g}=2.4%
\protect\nu $, so that $\protect\eta _{g}-\protect\eta _{\Omega }=2\protect%
\nu $. Here $t_{\ast }=24440\protect\nu ^{-1}$.}
\label{fig7}
\end{figure}

Now we consider the second-order transitions that take place during the
simultaneous $\Omega $- and $g$-modulations. In Fig. \ref{fig6} we set the
parameters $\varepsilon _{\Omega }=0.02\Omega _{0}$, $\varepsilon
_{g}=-0.02g_{0}$, $\eta _{\Omega }=0.4\nu $ and $\eta _{g}=1.6\nu $ (with
other parameters unaltered) in order to induce the transition $|\varphi
_{n}^{(1,2)}\rangle \rightarrow |\varphi _{n+2}^{(1,2)}\rangle $ via the
resonance $\eta _{g}+\eta _{\Omega }=2\nu =r$. In Fig. \ref{fig7} we use the
same parameters, except for $\eta _{g}=2.4\nu $, to induce to the
resonance $\eta _{g}-\eta _{\Omega }=2\nu $. In both cases the dynamics is
qualitatively similar to the behavior observed in Fig. \ref{fig4}.

\begin{figure}[tbh!]
\begin{center}
\includegraphics[width=0.65\textwidth]{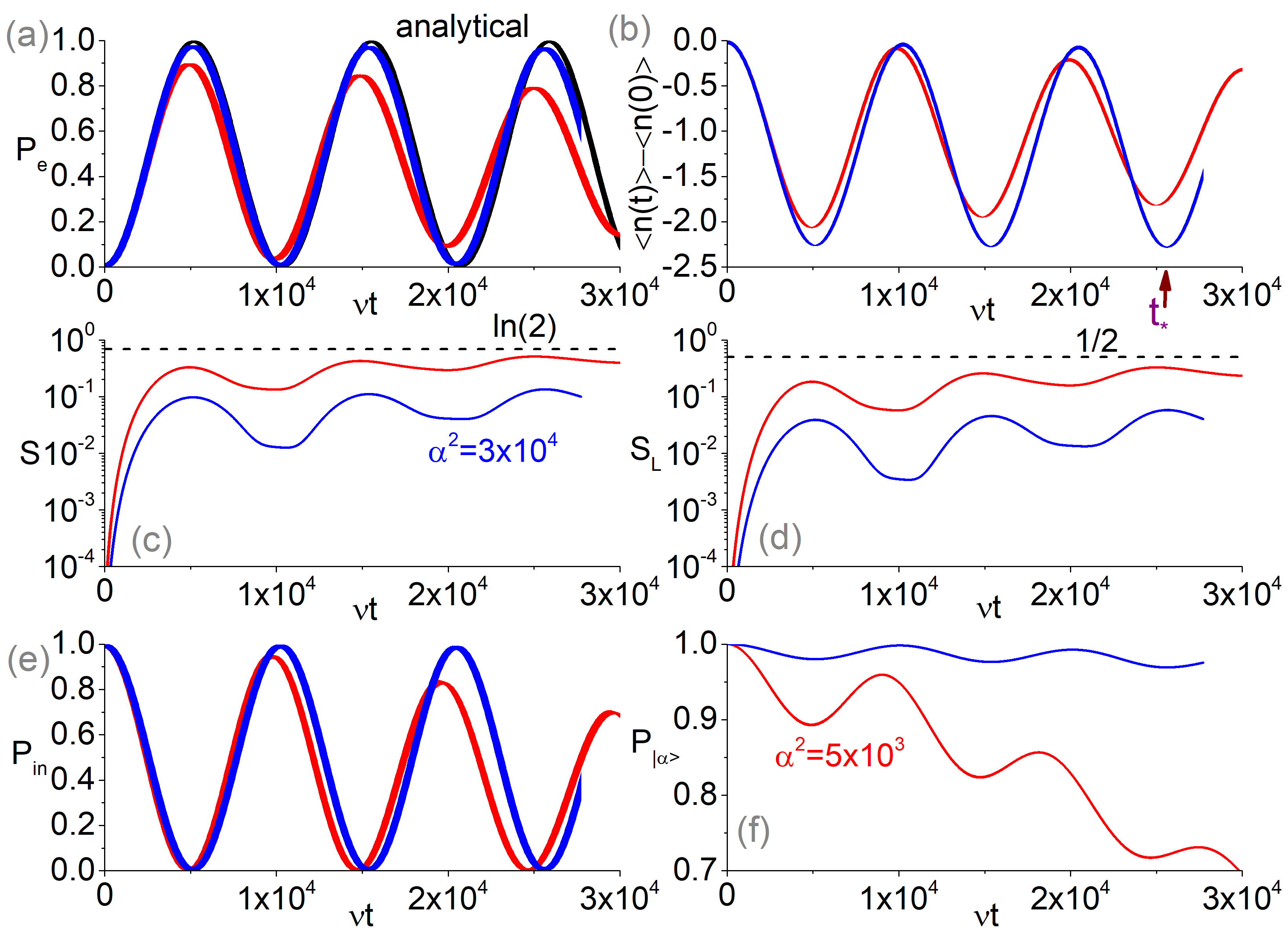} {}
\end{center}
\caption{Similar to Fig. \protect\ref{fig6} but for transitions $|\protect%
\varphi _{n}\rangle \rightarrow |\protect\varphi _{n+4}\rangle $, where $%
\protect\varphi $ stands for $S$ and $A$. Parameters: $\protect\varepsilon %
_{\Omega }=0.04\Omega _{0}$, $\protect\varepsilon _{g}=-0.04g_{0}$, $\protect%
\eta _{\Omega }=2.3\protect\nu $ and $\protect\eta _{g}=1.7\protect\nu $, so
that $\protect\eta _{g}+\protect\eta _{\Omega }=4\protect\nu $. Here $%
t_{\ast }=25170\protect\nu ^{-1}$.}
\label{fig8}
\end{figure}

\begin{figure}[tbh!]
\begin{center}
\includegraphics[width=0.65\textwidth]{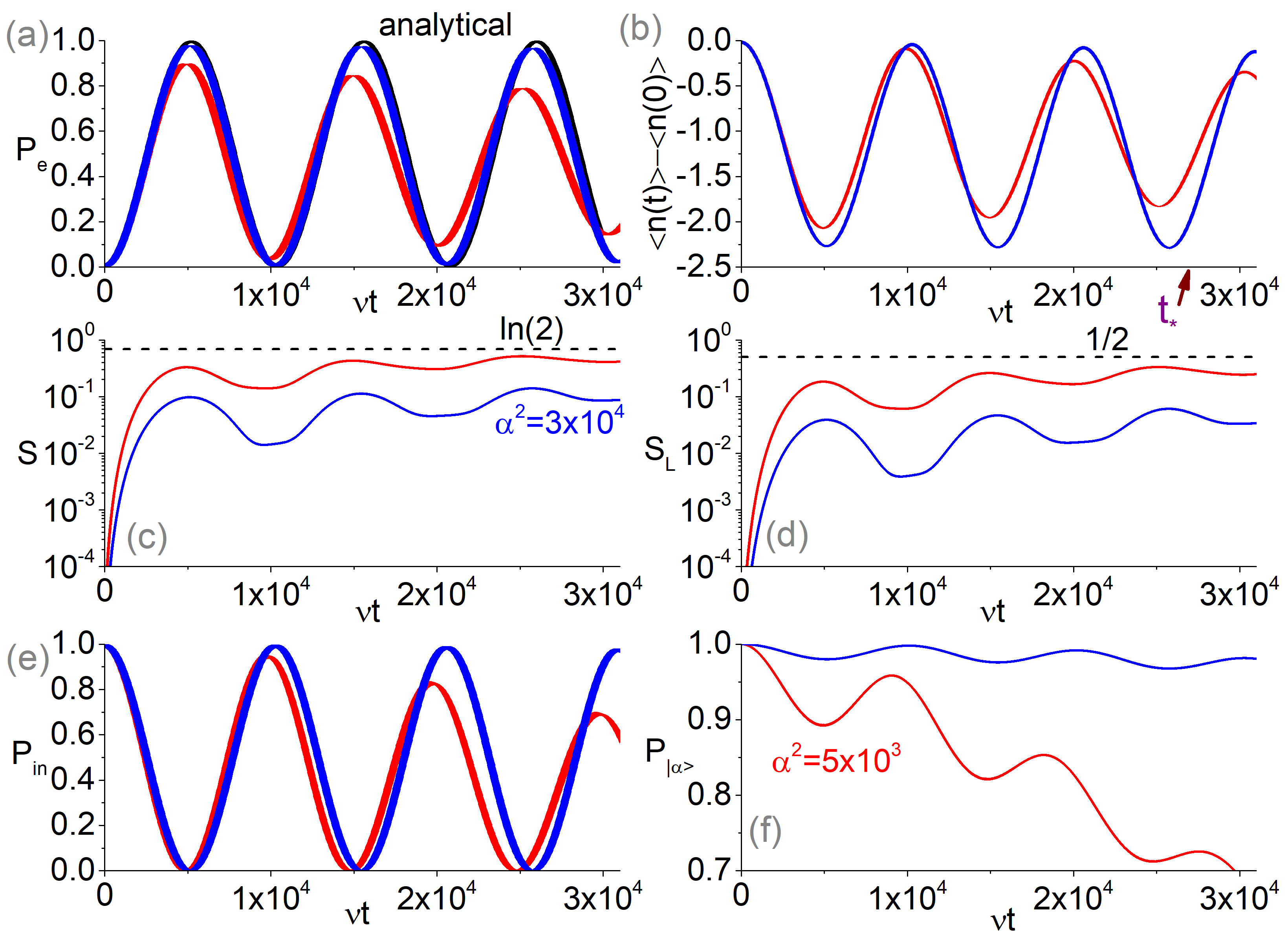} {}
\end{center}
\caption{Similar to Fig. \protect\ref{fig8} but for $\protect\eta _{g}=6.3%
\protect\nu $, so that $\protect\eta _{g}-\protect\eta _{\Omega }=4\protect%
\nu $. Here $t_{\ast }=26950\protect\nu ^{-1}$.}
\label{fig9}
\end{figure}

Finally, in Figs. \ref{fig8} and \ref{fig9} we set the parameters $%
\varepsilon _{\Omega }=0.04\Omega _{0}$ and $\varepsilon _{g}=-0.04g_{0}$ to
induce the weaker transitions $|\varphi _{n}^{(1,2)}\rangle \rightarrow |\varphi
_{n+4}^{(1,2)}\rangle $ (other parameters are unaltered). In Fig. \ref{fig8} we set $\eta _{\Omega }=2.3\nu $
and $\eta _{g}=1.7\nu $, while in Fig. \ref{fig9}, $\eta _{\Omega }=2.3\nu $
and $\eta _{g}=6.3\nu $, in order to match the resonances $\eta _{g}\pm \eta
_{\Omega }=4\nu =2\nu +r$. The behaviors observed in these two figures are practically the same, apart
from a small difference in the total transition rate $\Xi $ in Eqs. (\ref{f1}%
)--(\ref{f2}).

Notice that, in all the cases considered here, the approximate analytic solutions for $P_e$ according to the SRM are in excellent agreement with the predictions of the QRM, as long as
$\alpha^2\gtrsim 3\times 10^4$ and one considers initial times.

\section{Conclusions\label{conc}}

In this paper we considered the lossless semiclassical and quantum Rabi
models in the regime of multiphoton resonances and external harmonic perturbation of atomic parameters. We considered the modulation
of the atomic transition frequency, atom--field coupling strength or the
simultaneous modulation of both. In particular, we studied the less common
``second-order resonances'', when the sum or difference of different
modulation frequencies match some resonance condition of the total system.

We obtained approximate analytical solutions for the SRM, which exhibited an
excellent agreement with exact numeric results. For the QRM, we
theoretically explained how the external modulation modifies the system
dynamics near multiphoton resonances for coherent states with large
amplitudes, by analyzing the transition rates among the dressed-states.

As the main result of this work, we solved numerically the QRM for the initial
coherent states with large average photon numbers ($5\times 10^3$ and $3\times 10^4$), and illustrated
how the behavior of the qubit population according to the QRM approaches the
semiclassical one as the amplitude of the coherent state increases. However,
for large times the predictions of QRM and SRM inevitably begin to differ
qualitatively due to the characteristic collapse--revival phenomenon in the
quantum model.

Finally, we illustrated how the cavity field is modified due to the
interaction with the atom. We showed the behavior of the average photon
number, von Neumann and linear entropies (which quantify the degree of
entanglement), probabilities of finding the system and the cavity field alone in
the original states, and the deviation of the photon number probability
distribution from the initial one. We showed that after a complete
oscillation of the atomic excitation probability, the cavity field does not
return to the previous state, and the atom--field entanglement grows as
function of time, attaining the maximum allowed value for large times.

Thus, we believe this work contributes to a better understanding of the actual dynamics that takes place
during the interaction of an atom with a strong (in other words,
``classical'') monochromatic coherent field. Although the semiclassical approach describes
well the qubit dynamics for initial times, the quantum treatment shows
explicitly how the photons are absorbed and reemitted, and how the state of
the cavity field is continuously modified due to the light--matter
entanglement. Finally, our results show that the semiclassical model
inevitably fails after some characteristic time interval, which depends on the
initial average number of photons and the atom--field coupling parameter.

\appendix
\setcounter{section}{0}
\section{Transition rates in SRM\label{appen}}

\subsection{First-order coefficients in Eq. (\protect\ref{cof1})\label%
{appendixA}}

\[
\frac{e^{-i\phi }\Xi _{-r,\eta _{\Omega }}}{i}=\left( \frac{\tilde{g}%
_{0}\varepsilon _{\Omega }}{2}J_{0}^{0}J_{0}^{1}+g_{0}J_{1}^{0}J_{1}^{1}%
\right) J_{0}^{2}J_{0}^{3}J_{0}^{4}
\]%
\begin{eqnarray*}
\frac{e^{-i\phi }\Xi _{-r,\eta _{g}}}{i} &=&\frac{\tilde{\Delta}\varepsilon
_{g}}{2}J_{0}^{0}J_{0}^{1}J_{0}^{2}J_{0}^{3}J_{0}^{4}+\tilde{R}%
_{-}g_{0}J_{0}^{0}J_{0}^{1}J_{0}^{2}J_{1}^{3}J_{0}^{4}-\tilde{R}%
_{+}g_{0}J_{0}^{0}J_{0}^{1}J_{0}^{2}J_{0}^{3}J_{1}^{4} +\frac{\varepsilon _{g}}{2}%
J_{1}^{0}J_{0}^{1}J_{0}^{2}J_{0}^{3}J_{0}^{4}\\
&&-g_{0}J_{1}^{0}J_{0}^{1}J_{1}^{2}J_{0}^{3}J_{0}^{4}
+\tilde{R}_{-}g_{0}J_{2}^{0}J_{0}^{1}J_{0}^{2}J_{0}^{3}J_{1}^{4}-\tilde{R}%
_{+}g_{0}J_{2}^{0}J_{0}^{1}J_{0}^{2}J_{1}^{3}J_{0}^{4}
\end{eqnarray*}%
\[
\frac{e^{-i\phi }\Xi _{2-r,\eta _{\Omega }}}{i}=\tilde{R}%
_{+}g_{0}J_{0}^{0}J_{1}^{1}J_{0}^{2}J_{0}^{3}J_{0}^{4}-\tilde{R}%
_{-}g_{0}J_{2}^{0}J_{1}^{1}J_{0}^{2}J_{0}^{3}J_{0}^{4}-\frac{\tilde{g}%
_{0}\varepsilon _{\Omega }}{2}J_{1}^{0}J_{0}^{1}J_{0}^{2}J_{0}^{3}J_{0}^{4}
\]%
\begin{eqnarray*}
\frac{e^{-i\phi }\Xi _{2-r,\eta _{g}}}{i} &=&\frac{\tilde{R}_{+}\varepsilon
_{g}}{2}J_{0}^{0}J_{0}^{1}J_{0}^{2}J_{0}^{3}J_{0}^{4}-\tilde{R}%
_{+}g_{0}J_{0}^{0}J_{0}^{1}J_{1}^{2}J_{0}^{3}J_{0}^{4}-\frac{\tilde{\Delta}%
\varepsilon _{g}}{2}J_{1}^{0}J_{0}^{1}J_{0}^{2}J_{0}^{3}J_{0}^{4} \\
&&-\tilde{R}_{-}g_{0}J_{1}^{0}J_{0}^{1}J_{0}^{2}J_{1}^{3}J_{0}^{4}+\tilde{R}%
_{+}g_{0}J_{1}^{0}J_{0}^{1}J_{0}^{2}J_{0}^{3}J_{1}^{4}-\tilde{R}%
_{+}g_{0}J_{1}^{0}J_{0}^{1}J_{0}^{2}J_{1}^{3}J_{0}^{4} \\
&&-\frac{\tilde{R}_{-}\varepsilon _{g}}{2}%
J_{2}^{0}J_{0}^{1}J_{0}^{2}J_{0}^{3}J_{0}^{4}+\tilde{R}%
_{-}g_{0}J_{2}^{0}J_{0}^{1}J_{1}^{2}J_{0}^{3}J_{0}^{4}-\tilde{R}%
_{-}g_{0}J_{3}^{0}J_{0}^{1}J_{0}^{2}J_{0}^{3}J_{1}^{4}
\end{eqnarray*}

\[
\frac{e^{-i\phi }\Xi _{2-r,-\eta _{\Omega }}}{i}=\tilde{R}%
_{+}g_{0}J_{0}^{0}J_{1}^{1}J_{0}^{2}J_{0}^{3}J_{0}^{4}+\frac{\tilde{g}%
_{0}\varepsilon _{\Omega }}{2}J_{1}^{0}J_{0}^{1}J_{0}^{2}J_{0}^{3}J_{0}^{4}-%
\tilde{R}_{-}g_{0}J_{2}^{0}J_{1}^{1}J_{0}^{2}J_{0}^{3}J_{0}^{4}
\]%
\begin{eqnarray*}
\frac{e^{-i\phi }\Xi _{2-r,-\eta _{g}}}{i} &=&-\frac{\tilde{R}%
_{+}\varepsilon _{g}}{2}J_{0}^{0}J_{0}^{1}J_{0}^{2}J_{0}^{3}J_{0}^{4}-\tilde{%
R}_{+}g_{0}J_{0}^{0}J_{0}^{1}J_{1}^{2}J_{0}^{3}J_{0}^{4}+\frac{\tilde{\Delta}%
\varepsilon _{g}}{2}J_{1}^{0}J_{0}^{1}J_{0}^{2}J_{0}^{3}J_{0}^{4} \\
&&-\tilde{R}_{-}g_{0}J_{1}^{0}J_{0}^{1}J_{0}^{2}J_{0}^{3}J_{1}^{4}+\tilde{R}%
_{+}g_{0}J_{1}^{0}J_{0}^{1}J_{0}^{2}J_{1}^{3}J_{0}^{4}-\tilde{R}%
_{+}g_{0}J_{1}^{0}J_{0}^{1}J_{0}^{2}J_{0}^{3}J_{1}^{4} \\
&&+\frac{\tilde{R}_{-}\varepsilon _{g}}{2}%
J_{2}^{0}J_{0}^{1}J_{0}^{2}J_{0}^{3}J_{0}^{4}+\tilde{R}%
_{-}g_{0}J_{2}^{0}J_{0}^{1}J_{1}^{2}J_{0}^{3}J_{0}^{4}-\tilde{R}%
_{-}g_{0}J_{3}^{0}J_{0}^{1}J_{0}^{2}J_{1}^{3}J_{0}^{4}
\end{eqnarray*}%
\[
\frac{e^{-i\phi }\Xi _{-2-r,\eta _{\Omega }}}{i}=-\tilde{R}%
_{-}g_{0}J_{0}^{0}J_{1}^{1}J_{0}^{2}J_{0}^{3}J_{0}^{4}+\frac{\tilde{g}%
_{0}\varepsilon _{\Omega }}{2}J_{1}^{0}J_{0}^{1}J_{0}^{2}J_{0}^{3}J_{0}^{4}+%
\tilde{R}_{+}g_{0}J_{2}^{0}J_{1}^{1}J_{0}^{2}J_{0}^{3}J_{0}^{4}
\]%
\begin{eqnarray*}
\frac{e^{-i\phi }\Xi _{-2-r,\eta _{g}}}{i} &=&\frac{\varepsilon _{g}}{2}%
J_{0}^{1}J_{0}^{2}J_{0}^{3}J_{0}^{4}\left( \tilde{R}_{+}J_{2}^{0}-\tilde{R}%
_{-}J_{0}^{0}+\tilde{\Delta}J_{1}^{0}\right) \\
&&+\tilde{R}_{-}g_{0}J_{0}^{1}\left(
J_{0}^{0}J_{1}^{2}J_{0}^{3}J_{0}^{4}-J_{1}^{0}J_{0}^{2}J_{0}^{3}J_{1}^{4}+J_{1}^{0}J_{0}^{2}J_{1}^{3}J_{0}^{4}\right)
\\
&&-\tilde{R}_{+}g_{0}J_{0}^{1}\left(
J_{1}^{0}J_{0}^{2}J_{0}^{3}J_{1}^{4}+J_{2}^{0}J_{1}^{2}J_{0}^{3}J_{0}^{4}+J_{3}^{0}J_{0}^{2}J_{1}^{3}J_{0}^{4}\right)
\end{eqnarray*}%
\[
\frac{e^{-i\phi }\Xi _{4-r,\eta _{\Omega }}}{i}=-\tilde{R}%
_{+}g_{0}J_{1}^{0}J_{1}^{1}J_{0}^{2}J_{0}^{3}J_{0}^{4}+\frac{\tilde{g}%
_{0}\varepsilon _{\Omega }}{2}J_{2}^{0}J_{0}^{1}J_{0}^{2}J_{0}^{3}J_{0}^{4}+%
\tilde{R}_{-}g_{0}J_{3}^{0}J_{1}^{1}J_{0}^{2}J_{0}^{3}J_{0}^{4}
\]%
\begin{eqnarray*}
\frac{e^{-i\phi }\Xi _{4-r,\eta _{g}}}{i} &=&-\tilde{R}%
_{+}g_{0}J_{0}^{0}J_{0}^{1}J_{0}^{2}J_{1}^{3}J_{0}^{4}-\frac{\tilde{R}%
_{+}\varepsilon _{g}}{2}J_{1}^{0}J_{0}^{1}J_{0}^{2}J_{0}^{3}J_{0}^{4}+\tilde{%
R}_{+}g_{0}J_{1}^{0}J_{0}^{1}J_{1}^{2}J_{0}^{3}J_{0}^{4} \\
&&+\frac{\tilde{\Delta}\varepsilon _{g}}{2}%
J_{2}^{0}J_{0}^{1}J_{0}^{2}J_{0}^{3}J_{0}^{4}+\tilde{R}%
_{-}g_{0}J_{2}^{0}J_{0}^{1}J_{0}^{2}J_{1}^{3}J_{0}^{4}-\tilde{R}%
_{+}g_{0}J_{2}^{0}J_{0}^{1}J_{0}^{2}J_{0}^{3}J_{1}^{4} \\
&&+\frac{\tilde{R}_{-}\varepsilon _{g}}{2}%
J_{3}^{0}J_{0}^{1}J_{0}^{2}J_{0}^{3}J_{0}^{4}-\tilde{R}%
_{-}g_{0}J_{3}^{0}J_{0}^{1}J_{1}^{2}J_{0}^{3}J_{0}^{4}+\tilde{R}%
_{+}g_{0}J_{3}^{0}J_{0}^{1}J_{1}^{2}J_{0}^{3}J_{0}^{4}
\end{eqnarray*}%
\[
\frac{e^{-i\phi }\Xi _{4-r,-\eta _{\Omega }}}{i}=-\tilde{R}%
_{+}g_{0}J_{1}^{0}J_{1}^{1}J_{0}^{2}J_{0}^{3}J_{0}^{4}-\frac{\tilde{g}%
_{0}\varepsilon _{\Omega }}{2}J_{2}^{0}J_{0}^{1}J_{0}^{2}J_{0}^{3}J_{0}^{4}+%
\tilde{R}_{-}g_{0}J_{3}^{0}J_{1}^{1}J_{0}^{2}J_{0}^{3}J_{0}^{4}
\]%
\begin{eqnarray*}
\frac{e^{-i\phi }\Xi _{4-r,-\eta _{g}}}{i} &=&\frac{\tilde{R}%
_{+}\varepsilon _{g}}{2}J_{1}^{0}J_{0}^{1}J_{0}^{2}J_{0}^{3}J_{0}^{4}-\tilde{R}%
_{+}g_{0}J_{0}^{0}J_{0}^{1}J_{0}^{2}J_{0}^{3}J_{1}^{4}+\tilde{%
R}_{+}g_{0}J_{1}^{0}J_{0}^{1}J_{1}^{2}J_{0}^{3}J_{0}^{4}-\frac{\tilde{\Delta}\varepsilon _{g}}{2}%
J_{2}^{0}J_{0}^{1}J_{0}^{2}J_{0}^{3}J_{0}^{4} \\
&+&\tilde{R}%
_{-}g_{0}J_{2}^{0}J_{0}^{1}J_{0}^{2}J_{0}^{3}J_{1}^{4}-\tilde{R}%
_{+}g_{0}J_{2}^{0}J_{0}^{1}J_{0}^{2}J_{1}^{3}J_{0}^{4} -\frac{\tilde{R}_{-}\varepsilon _{g}}{2}%
J_{3}^{0}J_{0}^{1}J_{0}^{2}J_{0}^{3}J_{0}^{4}-\tilde{R}%
_{-}g_{0}J_{3}^{0}J_{0}^{1}J_{1}^{2}J_{0}^{3}J_{0}^{4}
\end{eqnarray*}%
\[
\frac{e^{-i\phi }\Xi _{-4-r,\eta _{\Omega }}}{i}=-\tilde{R}%
_{-}g_{0}J_{1}^{0}J_{1}^{1}J_{0}^{2}J_{0}^{3}J_{0}^{4}+\frac{\tilde{g}%
_{0}\varepsilon _{\Omega }}{2}J_{2}^{0}J_{0}^{1}J_{0}^{2}J_{0}^{3}J_{0}^{4}+%
\tilde{R}_{+}g_{0}J_{3}^{0}J_{1}^{1}J_{0}^{2}J_{0}^{3}J_{0}^{4}
\]%
\begin{eqnarray*}
\frac{e^{-i\phi }\Xi _{-4-r,\eta _{g}}}{i} &=&\tilde{R}%
_{-}g_{0}J_{0}^{0}J_{0}^{1}J_{0}^{2}J_{0}^{3}J_{1}^{4}-\frac{\tilde{R}%
_{-}\varepsilon _{g}}{2}J_{1}^{0}J_{0}^{1}J_{0}^{2}J_{0}^{3}J_{0}^{4}+\tilde{%
R}_{-}g_{0}J_{1}^{0}J_{0}^{1}J_{1}^{2}J_{0}^{3}J_{0}^{4} +\frac{\tilde{\Delta}\varepsilon _{g}}{2}%
J_{2}^{0}J_{0}^{1}J_{0}^{2}J_{0}^{3}J_{0}^{4}\\
&&+\tilde{R}%
_{-}g_{0}J_{2}^{0}J_{0}^{1}J_{0}^{2}J_{1}^{3}J_{0}^{4}-\tilde{R}%
_{+}g_{0}J_{2}^{0}J_{0}^{1}J_{0}^{2}J_{0}^{3}J_{1}^{4} +\frac{\tilde{R}_{+}\varepsilon _{g}}{2}%
J_{3}^{0}J_{0}^{1}J_{0}^{2}J_{0}^{3}J_{0}^{4}-\tilde{R}%
_{+}g_{0}J_{3}^{0}J_{0}^{1}J_{1}^{2}J_{0}^{3}J_{0}^{4}
\end{eqnarray*}%
\[
\frac{e^{-i\phi }\Xi _{6-r,\eta _{\Omega }}}{i}=\tilde{R}%
_{+}g_{0}J_{2}^{0}J_{1}^{1}J_{0}^{2}J_{0}^{3}J_{0}^{4}-\frac{\tilde{g}%
_{0}\varepsilon _{\Omega }}{2}J_{3}^{0}J_{0}^{1}J_{0}^{2}J_{0}^{3}J_{0}^{4}
\]%
\begin{eqnarray*}
\frac{e^{-i\phi }\Xi _{6-r,\eta _{g}}}{i} &=&\tilde{R}%
_{+}g_{0}J_{1}^{0}J_{0}^{1}J_{0}^{2}J_{1}^{3}J_{0}^{4}+\frac{\tilde{R}%
_{+}\varepsilon _{g}}{2}J_{2}^{0}J_{0}^{1}J_{0}^{2}J_{0}^{3}J_{0}^{4}-\tilde{%
R}_{+}g_{0}J_{2}^{0}J_{0}^{1}J_{1}^{2}J_{0}^{3}J_{0}^{4} \\
&&-\frac{\tilde{\Delta}\varepsilon _{g}}{2}%
J_{3}^{0}J_{0}^{1}J_{0}^{2}J_{0}^{3}J_{0}^{4}-\tilde{R}%
_{-}g_{0}J_{3}^{0}J_{0}^{1}J_{0}^{2}J_{1}^{3}J_{0}^{4}+\tilde{R}%
_{+}g_{0}J_{3}^{0}J_{0}^{1}J_{0}^{2}J_{0}^{3}J_{1}^{4}
\end{eqnarray*}%
\[
\frac{e^{-i\phi }\Xi _{6-r,-\eta _{\Omega }}}{i}=\tilde{R}%
_{+}g_{0}J_{2}^{0}J_{1}^{1}J_{0}^{2}J_{0}^{3}J_{0}^{4}+\frac{\tilde{g}%
_{0}\varepsilon _{\Omega }}{2}J_{3}^{0}J_{0}^{1}J_{0}^{2}J_{0}^{3}J_{0}^{4}
\]%
\begin{eqnarray*}
\frac{e^{-i\phi }\Xi _{6-r,-\eta _{g}}}{i} &=&\tilde{R}%
_{+}g_{0}J_{1}^{0}J_{0}^{1}J_{0}^{2}J_{0}^{3}J_{1}^{4}-\frac{\tilde{R}%
_{+}\varepsilon _{g}}{2}J_{2}^{0}J_{0}^{1}J_{0}^{2}J_{0}^{3}J_{0}^{4}-\tilde{%
R}_{+}g_{0}J_{2}^{0}J_{0}^{1}J_{1}^{2}J_{0}^{3}J_{0}^{4} \\
&&+\frac{\tilde{\Delta}\varepsilon _{g}}{2}%
J_{3}^{0}J_{0}^{1}J_{0}^{2}J_{0}^{3}J_{0}^{4}-\tilde{R}%
_{-}g_{0}J_{3}^{0}J_{0}^{1}J_{0}^{2}J_{0}^{3}J_{1}^{4}+\tilde{R}%
_{+}g_{0}J_{3}^{0}J_{0}^{1}J_{0}^{2}J_{1}^{3}J_{0}^{4}
\end{eqnarray*}%
\[
\frac{e^{-i\phi }\Xi _{-6-r,\eta _{\Omega }}}{i}=-\tilde{R}%
_{-}g_{0}J_{2}^{0}J_{1}^{1}J_{0}^{2}J_{0}^{3}J_{0}^{4}+\frac{\tilde{g}%
_{0}\varepsilon _{\Omega }}{2}J_{3}^{0}J_{0}^{1}J_{0}^{2}J_{0}^{3}J_{0}^{4}
\]%
\begin{eqnarray*}
\frac{e^{-i\phi }\Xi _{-6-r+\eta _{g}}}{i} &=&\tilde{R}%
_{-}g_{0}J_{1}^{0}J_{0}^{1}J_{0}^{2}J_{0}^{3}J_{1}^{4}-\frac{\tilde{R}%
_{-}\varepsilon _{g}}{2}J_{2}^{0}J_{0}^{1}J_{0}^{2}J_{0}^{3}J_{0}^{4}+\tilde{%
R}_{-}g_{0}J_{2}^{0}J_{0}^{1}J_{1}^{2}J_{0}^{3}J_{0}^{4} \\
&&+\frac{\tilde{\Delta}\varepsilon _{g}}{2}%
J_{3}^{0}J_{0}^{1}J_{0}^{2}J_{0}^{3}J_{0}^{4}+\tilde{R}%
_{-}g_{0}J_{3}^{0}J_{0}^{1}J_{0}^{2}J_{1}^{3}J_{0}^{4}-\tilde{R}%
_{+}g_{0}J_{3}^{0}J_{0}^{1}J_{0}^{2}J_{0}^{3}J_{1}^{4}
\end{eqnarray*}

\subsection{Second-order coefficients in Eq. (\protect\ref{cof2})\label%
{appendixB}}

\begin{eqnarray*}
e^{-i\phi }\Xi _{-r,\eta _{\Omega }+\eta _{g}} &=&\frac{\tilde{g}%
_{0}\varepsilon _{\Omega }}{2}J_{0}^{1}\left(
J_{1}^{0}J_{0}^{2}J_{0}^{3}J_{1}^{4}-J_{0}^{0}J_{1}^{2}J_{0}^{3}J_{0}^{4}-J_{1}^{0}J_{0}^{2}J_{1}^{3}J_{0}^{4}\right)
\\
&&+\frac{\varepsilon _{g}}{2}J_{1}^{1}J_{0}^{2}J_{0}^{3}J_{0}^{4}\left(
\tilde{\Delta}J_{0}^{0}+J_{1}^{0}\right) +\tilde{R}%
_{-}g_{0}J_{1}^{1}J_{0}^{2}\left(
J_{0}^{0}J_{1}^{3}J_{0}^{4}+J_{2}^{0}J_{0}^{3}J_{1}^{4}\right) \\
&&-\tilde{R}_{+}g_{0}J_{1}^{1}J_{0}^{2}\left(
J_{0}^{0}J_{0}^{3}J_{1}^{4}+J_{2}^{0}J_{1}^{3}J_{0}^{4}\right)
-g_{0}J_{1}^{0}J_{1}^{1}J_{1}^{2}J_{0}^{3}J_{0}^{4}
\end{eqnarray*}%
\begin{eqnarray*}
e^{-i\phi }\Xi _{-r,\eta _{\Omega }-\eta _{g}} &=&\frac{\tilde{g}%
_{0}\varepsilon _{\Omega }}{2}J_{0}^{1}\left(
J_{1}^{0}J_{0}^{2}J_{1}^{3}J_{0}^{4}-J_{0}^{0}J_{1}^{2}J_{0}^{3}J_{0}^{4}-J_{1}^{0}J_{0}^{2}J_{0}^{3}J_{1}^{4}\right)
\\
&&-\frac{\varepsilon _{g}}{2}J_{1}^{1}J_{0}^{2}J_{0}^{3}J_{0}^{4}\left(
J_{1}^{0}+\tilde{\Delta}J_{0}^{0}\right) +\tilde{R}%
_{-}g_{0}J_{1}^{1}J_{0}^{2}\left(
J_{0}^{0}J_{0}^{3}J_{1}^{4}+J_{2}^{0}J_{1}^{3}J_{0}^{4}\right) \\
&&-\tilde{R}_{+}g_{0}J_{1}^{1}J_{0}^{2}\left(
J_{0}^{0}J_{1}^{3}J_{0}^{4}+J_{2}^{0}J_{0}^{3}J_{1}^{4}\right)
-g_{0}J_{1}^{0}J_{1}^{1}J_{1}^{2}J_{0}^{3}J_{0}^{4}
\end{eqnarray*}%
\begin{eqnarray*}
e^{-i\phi }\Xi _{-r,-\eta _{\Omega }+\eta _{g}} &=&\frac{\tilde{g}%
_{0}\varepsilon _{\Omega }}{2}J_{0}^{1}\left(
J_{0}^{0}J_{1}^{2}J_{0}^{3}J_{0}^{4}+J_{1}^{0}J_{0}^{2}J_{1}^{3}J_{0}^{4}-J_{1}^{0}J_{0}^{2}J_{0}^{3}J_{1}^{4}\right)
\\
&&+\frac{\varepsilon _{g}}{2}J_{1}^{1}J_{0}^{2}J_{0}^{3}J_{0}^{4}\left(
J_{1}^{0}+\tilde{\Delta}J_{0}^{0}\right) +\tilde{R}%
_{-}g_{0}J_{1}^{1}J_{0}^{2}\left(
J_{0}^{0}J_{1}^{3}J_{0}^{4}+J_{2}^{0}J_{0}^{3}J_{1}^{4}\right) \\
&&-\tilde{R}_{+}g_{0}J_{1}^{1}J_{0}^{2}\left(
J_{0}^{0}J_{0}^{3}J_{1}^{4}+J_{2}^{0}J_{1}^{3}J_{0}^{4}\right)
-g_{0}J_{1}^{0}J_{1}^{1}J_{1}^{2}J_{0}^{3}J_{0}^{4}
\end{eqnarray*}%
\begin{eqnarray*}
e^{-i\phi }\Xi _{-2-r,\eta _{\Omega }+\eta _{g}} &=&-\frac{\tilde{g}%
_{0}\varepsilon _{\Omega }}{2}J_{0}^{1}\left(
J_{0}^{0}J_{0}^{2}J_{0}^{3}J_{1}^{4}+J_{1}^{0}J_{1}^{2}J_{0}^{3}J_{0}^{4}+J_{2}^{0}J_{0}^{2}J_{1}^{3}J_{0}^{4}\right)
\\
&&+\frac{\varepsilon _{g}}{2}J_{1}^{1}J_{0}^{2}J_{0}^{3}J_{0}^{4}\left(
\tilde{R}_{+}J_{2}^{0}-\tilde{R}_{-}J_{0}^{0}+\tilde{\Delta}J_{1}^{0}\right)
\\
&&+\tilde{R}_{-}g_{0}J_{1}^{1}J_{0}^{4}\left(
J_{0}^{0}J_{1}^{2}J_{0}^{3}+J_{1}^{0}J_{0}^{2}J_{1}^{3}\right) \\
&&-\tilde{R}_{+}g_{0}J_{1}^{1}J_{0}^{4}\left(
J_{2}^{0}J_{1}^{2}J_{0}^{3}+J_{3}^{0}J_{0}^{2}J_{1}^{3}\right)
-g_{0}J_{1}^{0}J_{1}^{1}J_{0}^{2}J_{0}^{3}J_{1}^{4}
\end{eqnarray*}%
\begin{eqnarray*}
e^{-i\phi }\Xi _{-2-r,\eta _{\Omega }-\eta _{g}} &=&-\frac{\tilde{g}%
_{0}\varepsilon _{\Omega }}{2}J_{0}^{1}\left(
J_{0}^{0}J_{0}^{2}J_{1}^{3}J_{0}^{4}+J_{1}^{0}J_{1}^{2}J_{0}^{3}J_{0}^{4}+J_{2}^{0}J_{0}^{2}J_{0}^{3}J_{1}^{4}\right)
\\
&&-\frac{\varepsilon _{g}}{2}J_{1}^{1}J_{0}^{2}J_{0}^{3}J_{0}^{4}\left(
\tilde{R}_{+}J_{2}^{0}-\tilde{R}_{-}J_{0}^{0}+\tilde{\Delta}J_{1}^{0}\right)
\\
&&+\tilde{R}_{-}g_{0}J_{1}^{1}J_{0}^{3}\left(
J_{0}^{0}J_{1}^{2}J_{0}^{4}+J_{1}^{0}J_{0}^{2}J_{1}^{4}\right) \\
&&-\tilde{R}_{+}g_{0}J_{1}^{1}J_{0}^{3}\left(
J_{2}^{0}J_{1}^{2}J_{0}^{4}+J_{3}^{0}J_{0}^{2}J_{1}^{4}\right)
-g_{0}J_{1}^{0}J_{1}^{1}J_{0}^{2}J_{1}^{3}J_{0}^{4}
\end{eqnarray*}

\begin{eqnarray*}
e^{-i\phi }\Xi _{-2-r,-\eta _{\Omega }+\eta _{g}} &=&\frac{\tilde{g}%
_{0}\varepsilon _{\Omega }}{2}J_{0}^{1}\left(
J_{0}^{0}J_{0}^{2}J_{0}^{3}J_{1}^{4}+J_{1}^{0}J_{1}^{2}J_{0}^{3}J_{0}^{4}+J_{2}^{0}J_{0}^{2}J_{1}^{3}J_{0}^{4}\right)
\\
&&+\frac{\varepsilon _{g}}{2}J_{1}^{1}J_{0}^{2}J_{0}^{3}J_{0}^{4}\left(
\tilde{R}_{+}J_{2}^{0}-\tilde{R}_{-}J_{0}^{0}+\tilde{\Delta}J_{1}^{0}\right)
\\
&&+\tilde{R}_{-}g_{0}J_{1}^{1}J_{0}^{4}\left(
J_{0}^{0}J_{1}^{2}J_{0}^{3}+J_{1}^{0}J_{0}^{2}J_{1}^{3}\right) \\
&&-\tilde{R}_{+}g_{0}J_{1}^{1}J_{0}^{4}\left(
J_{2}^{0}J_{1}^{2}J_{0}^{3}+J_{3}^{0}J_{0}^{2}J_{1}^{3}\right)
-g_{0}J_{1}^{0}J_{1}^{1}J_{0}^{2}J_{0}^{3}J_{1}^{4}
\end{eqnarray*}

\section*{Acknowledgement}

M.V.S.P. thanks the Brazilian agency Coordenação de Aperfeiçoamento de Pessoal de Nível Superior (CAPES) for the financial support. A.V.D. acknowledges partial
financial support of the Brazilian agencies CNPq (Conselho Nacional de
Desenvolvimento Cient\'{\i}fico e Tecnol\'{o}gico) and Funda\c{c}\~{a}o de
Apoio \`{a} Pesquisa do Distrito Federal (FAPDF, grant number
00193-00001817/2023-43).

\newpage


\section*{References}

\begin{thebibliography}{99}
\bibitem{rabi1} Rabi, I.I. On the Process of Space Quantization. Phys. Rev. \textbf{1936}, 49, 324.

\bibitem{rabi2} Rabi, I. I. Space Quantization in a Gyrating Magnetic Field. Phys. Rev. \textbf{1937}, 51, 652.

\bibitem{bloch} Bloch, F.; Siegert. A. Magnetic Resonance for
Nonrotating Fields. Phys. Rev. \textbf{1940}, 57, 522.

\bibitem{irish} Irish, E.K.; Armour, A.D. Defining the Semiclassical Limit of
the Quantum Rabi Hamiltonian. Phys. Rev. Lett. \textbf{2022}, 129 183603.

\bibitem{shirley} Shirley, J.H. Solution of the Schr\"{o}dinger Equation
with a Hamiltonian Periodic in Time. Phys. Rev. \textbf{1965}, 138, B979.

\bibitem{duvall} Duvall, R.E.; Valeo, E.J.; Oberman, C.R. Nonperturbative
analysis of the two-level atom: Applications to multiphoton excitation. Phys. Rev. A \textbf{1988}, 37, 4685.

\bibitem{woerd} Beijersbergen, M.W.; Spreeuw, R.J.C.; Allen, L.;
Woerdman, J.P. Multiphoton resonances and Bloch-Siegert shifts observed in a
classical two-level system. Phys. Rev. A \textbf{1992}, 45, 1810.

\bibitem{ad7} Sainz, I.; Klimov, A.B.; Saavedra, C. Effective Hamiltonian
approach to periodically perturbed quantum optical systems. Phys. Lett. A \textbf{2006}, 351, 26.

\bibitem{castanos} Casta\~{n}os, L.O. Simple, analytic solutions of the
semiclassical Rabi model. Opt. Commun. \textbf{2019}, 430, 176.

\bibitem{saiko} Saiko, A.P.; Markevich, S.A.; Fedaruk, R. Bloch-Siegert oscillations in the Rabi model with an amplitude-modulated driving field. Laser Phys. \textbf{2019}, 29, 124004.

\bibitem{ad3} Sainz, I.; Garc\'{\i}a, A.; Klimov, A.B. Effective and
efficient resonant transitions in periodically modulated quantum systems. Quantum Reports \textbf{2021}, 3, 1.

\bibitem{simse} Chalkopiadis, L.; Simserides, C. Averaging method and
coherence applied to Rabi oscillations in a two-level system. J. Phys.
Commun. \textbf{2021}, 5, 095006

\bibitem{marinho0} Marinho, A.; Dodonov, A.V.
Approximate analytic solution of the dissipative semiclassical Rabi model under parametric multi-tone modulations.
Phys. Scr. \textbf{2024}, 99, 125117.

\bibitem{marinho1} Marinho A.; Dodonov, A. Analytic approach for dissipative semiclassical Rabi model under parametric modulation.
In \emph{Proceedings of the Second
International Workshop on Quantum Nonstationary Systems}; Dodonov A.; Ribeiro, C.C.H., Eds.; LF
Editorial: S\~{a}o Paulo, 2024; pp. 195-210 . DOI: 10.29327/5559154.1-12.

\bibitem{marinho2} Marinho, A.; de Paula, M.V.S.; Dodonov, A.V.
Approximate analytic solution of the dissipative semiclassical Rabi model
near the three-photon resonance and comparison with the quantum behavior.
Phys. Lett. A \textbf{2024}, 513, 129608.

\bibitem{boyd} Boyd, R.W. \emph{Nonlinear Optics}; Academic Press: London, 2nd Edition, 2003.

\bibitem{scully} Scully, M.O.; Zubairy, M.S. \emph{Quantum Optics};(Cambridge University Press: Cambridge, 1997.

\bibitem{shore} Shore, B.W. Coherent manipulation of atoms using laser
light. Acta Physica Slovaca \textbf{2008}, 58, 243.

\bibitem{luo} Yan, Y.; L\"{u},Z.; Luo, J.Y. Variational approach to
light-matter interaction: Bridging quantum and semiclassical limits. Phys.
Rev. A \textbf{2024}, 110, 013706.

\bibitem{acosta1} Costa A.P.; Dodonov, A. Quantum Rabi oscillations in
the semiclassical limit: backreaction on the cavity field and entanglement.
In \emph{Proceedings -- QNS III
International Workshop on Quantum Nonstationary Systems}; Dodonov A.; C\'{e}leri, L.C., Eds.; LF
Editorial: S\~{a}o Paulo, 2025; pp. 71-88 . DOI: 10.29327/5559166.

\bibitem{braak} Braak, D. Integrability of the Rabi model. Phys. Rev. Lett. \textbf{2011}, 107, 100401.

\bibitem{rev} Xie, Q.; Zhong, H.; Batchelor, M.T.; Lee, C. The quantum Rabi
model: solution and dynamics. J. Phys. A.: Math. Theor. \textbf{2017}, 50, 113001.

\bibitem{larson} Larson J.; Mavrogordatos, Th. K. \emph{The
Jaynes-Cummings Model and Its Descendants}; IOP Publishing: Bristol, 2021.
https://arxiv.org/abs/2202.00330.

\bibitem{klim0} Klimov, A.B.; Sainz, I.; Chumakov, S.M. Resonance expansion
versus the rotating-wave approximation. Phys. Rev. A \textbf{2003}, 68 063811.

\bibitem{j1} Dodonov, A.V. Photon creation from vacuum and interactions
engineering in nonstationary circuit QED. J. Phys.: Conf. Ser. \textbf{2009}, 161, 012029.

\bibitem{j3} De Liberato, S.; Gerace, D.; Carusotto, I.; Ciuti, C.
Extracavity quantum vacuum radiation from a single qubit. Phys. Rev. A \textbf{2009}, 80, 053810.

\bibitem{j2} Dodonov, A.V. Analytical description of nonstationary circuit
QED in the dressed-states basis. J. Phys. A: Math. Theor. \textbf{2014}, 47, 285303.

\bibitem{3fot1} Ma, K.K.W.; Law, C.K. Three-photon resonance and adiabatic
passage in the large-detuning Rabi model. Phys. Rev. A \textbf{2015}, 92, 023842.

\bibitem{garzi} Garziano, L.; Stassi, R.; Macr\`{\i}, V.; Kockum, A.F.;
Savasta, S.; Nori, F. Multiphoton quantum Rabi oscillations in ultrastrong
cavity QED. Phys. Rev. A \textbf{2015}, 92, 063830.

\bibitem{j4} Dodonov, A.V. Dynamical Casimir effect via four- and
five-photon transitions using a strongly detuned atom. Phys. Rev. A \textbf{2019}, 100, 032510.

\bibitem{ma} Ma, K.K.W. Multiphoton resonance and chiral transport in the
generalized Rabi model. Phys. Rev. A \textbf{2020}, 102, 053709.

\bibitem{3fot3} Cong, L.; Felicetti, S.; Casanova, J.; Lamata, L.; Solano, E.;
Arrazola, I. Selective interactions in the quantum Rabi model. Phys. Rev. A \textbf{2020}, 101, 032350.

\bibitem{shus2} Coleman H.F.A.; Twyeffort, E.K. Spectral and dynamical
validity of the rotating-wave approximation in the quantum and semiclassical
Rabi models. J. Opt. Soc. Am. B \textbf{2024}, 41, C188.

\bibitem{tom} Silva E.L.S.; Dodonov, A.V. Analytical comparison of
the first- and second-order resonances for implementation of the dynamical
Casimir effect in nonstationary circuit QED. J. Phys. A: Math. Theor. \textbf{2016}, 49, 495304.

\bibitem{messina} Dodonov, A.V.; Militello, B.; Napoli, A.; Messina, A.
Effective Landau-Zener transitions in the circuit dynamical Casimir effect
with time-varying  modulation frequency. Phys. Rev. A \textbf{2016}, 93, 052505.

\bibitem{eberly} Eberly, J.H.; Narozhny, N.B.; Sanchez-Mondragon, J.J. Periodic Spontaneous Collapse and Revival in a Simple Quantum Model. Phys. Rev. Lett. \textbf{1980}, 44, 1323.

\end{thebibliography}
\end{document}